\documentclass[%
	final,
	notitlepage,
	narroweqnarray,
	inline,
	twoside,
	]{ieee}

\usepackage{times}

\begin{document}

\title[Untappable key distribution system: \\a one-time-pad booster]{%
       Untappable key distribution system: \\a one-time-pad booster}

\author[BARBOSA\&GRAAF]{Geraldo A. Barbosa and Jeroen van de Graaf
\authorinfo{G.\,A.\,Barbosa, PhD,  is the CEO of QuantaSec -- Consulting and Projects in Physical Cryptography Ltd., Av. Portugal 1558, Belo Horizonte MG 31550-000 Brazil. Phone: $+$11\,55\,(31)\,3441--4121, e-mail: GeraldoABarbosa@gmail.com}
\authorinfo{J.\,van de\,Graaf, PhD, Professor, Instituto de Ci\^encias Exatas, Dept. de Ci\^encia da Computa\c c\~ao, Av. Ant\^onio Carlos 6.627, Belo Horizonte MG 31270-901 Brazil. E-mail: jvdg@dcc.ufmg.br}
}

\firstpage{1}
\maketitle
\newcommand{\be}{\begin{equation}}
\newcommand{\ee}{\end{equation}}
\newcommand{\bea}{\begin{eqnarray}}
\newcommand{\eea}{\end{eqnarray}}

\begin{abstract}
One-time-pad (OTP) encryption simply cannot be cracked, even by a quantum computer. The need of sharing in a secure way supplies of symmetric random keys turned the method almost obsolete as a standing-alone method for fast and large volume telecommunication. Basically, this secure sharing of keys and their renewal, once exhausted, had to be done through couriers, in a slow and costly process. This paper presents a solution for this problem providing a fast and unlimited renewal of secure keys: An {\em untappable key distribution system} is presented and detailed.
This  {\em fast} key distribution system utilizes two layers of confidentially protection: 1) Physical noise intrinsic to the optical channel that turn  the coded signals into {\em stealth} signals and 2) Privacy amplification using a bit pool of refreshed entropy run after run, to eliminate any residual information.
 The resulting level of security is rigorously calculated and demonstrates that the level of information an eavesdropper could obtain is completely negligible.
The  random bit sequences, fast and securely distributed,  can be used to encrypt text, data or voice.
\end{abstract}


\section{Introduction}

A {\em key distribution} system that uses the intrinsic light noise of an optical carrier  to forbid an attacker E (or Eve) to extract clean signals from the transmitted ones was described  in Refs.~\cite{barbosa1} and \cite{barbosa2}.
The basic characteristics of that system is that the legitimate users, A (or Alice) and B (or Bob), are {\em not}  affected in the same way as E by the channel's noise. This asymmetry is caused by a starting {\em information} shared by A and B but not by E -- it produces a measurement {\em advantage} for A and B over Eve: Signals buried under noise for Eve and clear signals for A or B.

This work stresses basic theoretical and practical aspects of that physical system (Section \ref{transmission and reception})  and enhances its security by an explicit privacy amplification protocol (PA) (Sections \ref{increasing} and \ref{PA}). The  security level due to these two protection layers, physical and computational, is calculated and discussed. The physical implementation will be presented in a following work.

The use of optical noise to secure {\em encryption} of signals in telecommunication channels was analyzed in \cite{alphaeta} (alpha-eta {\em encryption} system). This system was tested in secure networks in US, in land-air tests (Optix/NuCrypt), reached market applications (NuCrypt LCC) and produced independent developments in Japan \cite{hirota}. The system discussed here is akin to the alpha-eta (or $\alpha\eta$) system in the use of optical noise but is specific as a {\em key distribution} system.  Furthermore, it uses true random bit generators instead of linear feedback-shift-registers (LFSR) used in the alpha-eta systems.

 The resulting securely shared random sequences of bits can be used for encryption in arbitrary communication channels. It can be used for bit-to-bit encryption as a ``one-time pad'' system, with constantly renewed keys in a fast process.

The discussed system has a basic connection in principle with Ref.~\cite{grangier}, where cryptography using continuous variables with coherent states was proposed: the use of the {\em optical noise} is also at the core of that scheme.  However, several important differences exist between these systems. Among them, optical quantum demolition measurements are not necessary, neither quadrature measurements. Therefore, the  system  discussed in this paper is widely different and, as such, no need for comparisons exist.

The key distribution system presented here (as in \cite{barbosa1}) uses light's noise for protection by modulating each signal representing a random bit by  another randomly chosen (secret) signal representing a physical basis. The superposition of ``noise'' signals, ``basis modulation'' and ``bit'' signals frustrates any attacker trying to obtain either the basis used or the  transmitted bit. A privacy amplification protocol (PA) operates in a {\em bit-pool} constantly renewed in entropy that enhances the security of the system. The overall security achieved is calculated giving the users a guaranteed security level.

This system is designed such that its ultimate security should depend only on the secure transmission of the bits and their safe storage. In a sense, the system can be fully understood and signals openly acessed by the adversary and yet full security for A and B resides just on the keys (Kerckhoff's principle).

One among  the uncountable uses for this  continuously renewable and fast one time process,  is the protection of energy infrastructures (generation, distribution and their control interconnected by smart-grids).
It involves not only eventual private interests but above all it should rest as a main government interest - the development of a nation and the wealthy of the people depend on these infrastructures.

This paper is roughly divided in two parts,  one dealing with the  physical noise and other with privacy amplification aspects. Although the subjects are different, they are intrinsically connected by the architecture of the key distribution system and are essential for a full understanding of this system.

\section{Phase modulation and optical noise}
\label{PhaseModulation}

One of the simplest ways to implement the physical part of the scheme is using optical phase modulation of a laser beam. This modulation is achieved by fastly modifying the refractive index of an optical medium  in the beam path before the transmission channel (fiber optics or any continuous media).

{\em Bits} could be represented by a given phase, say $\phi_1=0^{\circ}$ for bit 1 and bit 0 by phase $\phi_0=180^{\circ}$. A {\em basis} is represented by an extra phase $\phi_b$, within a manyfold of $M$ possible values, that is added to  $\phi_1$ or $\phi_0$ producing a different resulting phase: $\phi_{b,j}=\phi_j+\phi_b, \: (j=1,0)$ and $(b=1,2,3 \dots M)$. Both bit and basis are unknown to the adversary.
\begin{figure}[h!]
\centerline{\scalebox{0.4}{\includegraphics{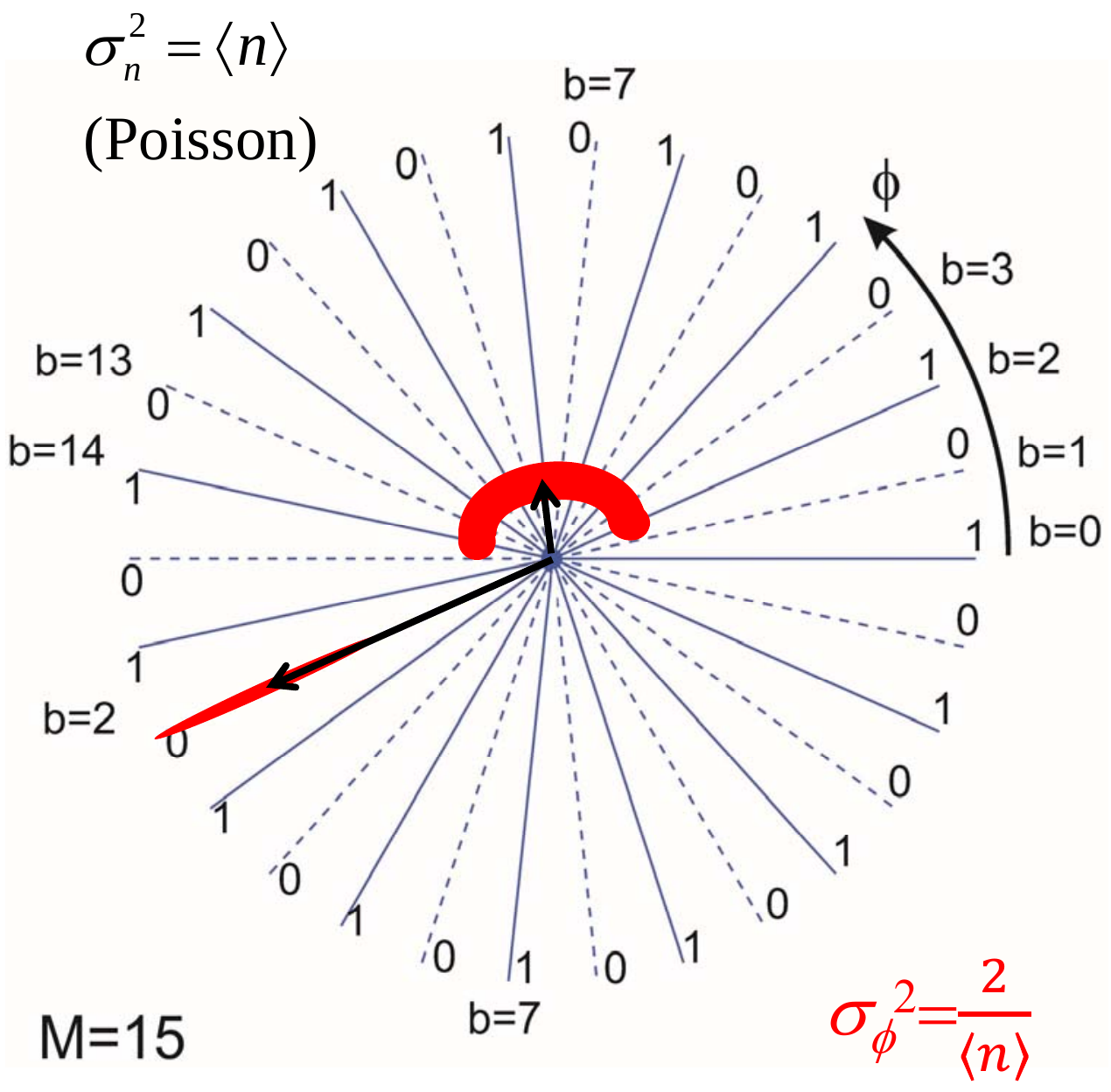}}}
\caption{ \label{BasesBitsAmplitudesPhases_3}  Wheel of phases representing encryption bases for bits. Bits 0 and 1 are represented at extremes of a basis and separated by $\pi$. Encryption bases are separated by $\pi/M$. A bit signal represented by an amplitude and phase has an intrinsic phase noise (given by $\sigma_{\phi}^2=2/\langle n \rangle$) that may cover adjacent bases and do not allows an attacker to identify the signal being sent. While a strong amplitude signal allows easy identification of the bases and bit (e.g., see signal at basis $b=2$), a weak signal does not allow such identification (e.g., see signal around basis 7 or 8). It should be emphasized that a signal representing one bit is sent just once and never repeated.}
\end{figure}
Fig.~\ref{BasesBitsAmplitudesPhases_3} explain these ideas. It represents a uniformly spaced set of bases constituted of $M$ physical phase bases ($M=15: b=0,1,2,\dots,14$ in this example, closest bases are separated by an angle of $12^{\circ}$).
Each basis is represented as a {\em single} line made up of a solid line continued  by a dashed one (a given phase value and this same value $+ 180^{\circ}$ represents a single basis).

The fundamental characteristics of these bases is that bits are represented in an alternate order in neighboring bases. For example, if Alice wants to send a bit 1 she may pick one basis, say $b=2$ (without Eve's knowledge). In this basis, bit 1 is represented along a solid line $(\phi_2=24^{\circ})$. If she had picked $b=3$, bit 1 would have been placed at the phase $\phi=(12\times 3+180)^{\circ}$. In any of these cases, the closest bits in neighboring bases would have been opposite.

The distance from the center, along any basis lines in Fig.~\ref{BasesBitsAmplitudesPhases_3}, gives the {\em amplitude} of the light field that carries the bit signal while phases are represented around a circle as indicated.  The uncertainty in the signal to be measured (e.g., by measuring Stokes parameters) can be represented in Fig.~\ref{BasesBitsAmplitudesPhases_3} by a smeared figure representing uncertainties in amplitude and phase (see red features in figure).

A difference between a strong and a weak coherent signal is that the phase uncertainty over the average signal level with $n$ photons ($1/\sqrt{\langle n \rangle}$) for the weak signal is larger.

The phase uncertainty can be calculated in a similar way as done in the polarization uncertainty obtained in \cite{barbosa1} (Eq.~(2)): Assume that a laser beam in a coherent state $|\Psi_0 \rangle=|\alpha \rangle$  passes through an optical modulator that produces a phase difference $\phi$ between its two physical axis (say $x$ and $y$) for an incoming polarization state. One should recall that for a coherent state $|\alpha|^2=\langle n \rangle$ \cite{Glauber}.

The optical modulator is a two-port device for an incoming state. The modulator transforms the state $|\alpha \rangle$ according to the angular momentum rotation operator $J_z$ for two states (or two modes). These states are represented by photon annihilation operators $a_x$ and $b_y$: $J_z=(1/2)(a_x^+ a_x-b_y^+ b_y)$ \cite{MessiahII}.   The transformation produces
\bea
\label{jz wavefunctions}
|\Psi(\phi) \rangle&=&e^{-i J_z \phi} |\Psi_0 \rangle \nonumber \\
& =&|\frac{\alpha}{\sqrt{2}} e^{-i \phi/2} \rangle_x   |\frac{\alpha}{\sqrt{2}} e^{i \phi/2} \rangle_y \:.
\eea
In $|\Psi(\phi) \rangle$ a phase is established due to a phase difference between two orthogonal components $x$ and $y$.
 From this result the overlap of two states $|\Psi(\phi) \rangle$ and $|\Psi(\phi') \rangle$ can be obtained:
\bea
\label{overlap}
\langle \Psi(\phi)|\Psi(\phi') \rangle=e^{-|\alpha|^2\left[1-\cos(\frac{\phi-\phi'}{2})  \right]}\:.
\eea
This overlap is a measure of the ``indistinguishability" degree between the two states.
For mesoscopic states $|\alpha|^2 \gg 1$ (but not intense) and the exponential term gives a vanishing contribution unless $\phi-\phi'$ is small. Considering $\Delta \phi\equiv \phi-\phi'\ll 1$ one has the Gaussian distribution
\bea
\langle \Psi(\phi)|\Psi(\phi') \rangle \rightarrow
e^{-|\alpha|^2 (\Delta \phi)^2/2} \:.
\eea

The probability for indistinguishability between $\phi$ and $\phi'$ is then given by
\bea
|\langle \Psi(\phi)|\Psi(\phi') \rangle|^2\rightarrow e^{-|\alpha|^2 (\Delta \phi)^2}\equiv
e^{   -(\Delta \phi)^2/(2 {\sigma_{\phi}}^2  )  }\:,
\eea
where $\sigma_{\phi}=\sqrt{2/ \langle n \rangle}$.

This shows that a strong signal (large $\langle n \rangle$) has a reduced phase uncertainty. For example, the large amplitude signal representing a bit 0 in basis $b=2$ could be easily identifiable (see Fig.~\ref{BasesBitsAmplitudesPhases_3}) by either A, B or E because the phase uncertainty is small and no confusion is possible with a neighboring bit.  Differently, if the phase uncertainty is such (weak signal) that the obtained signal overlaps neighboring bases (see uncertainty around basis $b=8$ in Fig.~\ref{BasesBitsAmplitudesPhases_3}), the information of {\em which} basis is being used is not available. Consequently, the bit sent cannot be identified without a large probability of error. This noisy channel can be referred as the $\alpha\eta$ channel.

However, if the legitimate users know which basis was used, there is no ambiguity in bit identification. For them, bit identification is just a question of identifying if the signal is {\em around} a given phase value $\phi$ or at $\phi+180^{\circ}$, not between closest bases where identification is not allowed due to the phase noise.

If the basis information is not available to Eve, her measurements will produce errors in the bases or bit estimation (for example, signals around bases 7 and 8). In other words, as the separation between closest bases is $\pi/M$, the resolution needed for bit or basis identification has to  be better than this value. However, the noise is tailored by the legitimate users to produce an uncertainty much larger than $\pi/M$ and the attacker has no way to reduce this noise.

With a proper choice of a separation $\Delta \phi$ between bases  and average number of photons $\langle n \rangle$, not only the separation can be set
$\Delta \phi < \sigma (=1/\sqrt{2 \langle n \rangle})$  but also the probability of an error by Eve, $P_e^{E}$ can be set arbitrarily close to 1/2 (see Fig. 3 in \cite{barbosa1} and discussions therein).
The derivation of $P_e$ using POVM (Positive Operator Valued Measurement) can be found in Ref.~\cite{barbosa1}.  Appendix~\ref{noisyPoincare} discuss the effect of noise in the channel in an alternate way, by means of Poincar\'e measurements. This way, the reader has different contexts for comparisons.

The different amount of information between E and the legitimate users produce the different results between E and B in a transmitted sequence of bits. This is made possible  by information shared beforehand in each transmission round by A and B  in the form of a secret stream with the information of the bases being used and about which E has no  information.

We will also see that this  process can be  continued without limitation,  without any need for A and B to meet after the first contact.

{\bf XOR encoding?--}
Some questions may be asked, such as ``why this bases encoding cannot be replaced by a simple  XOR of {\em basis}($=${\em single bit})  with the {\em bit}, especially in the model where the eavesdropper gets a perfect copy of the transmitted state?".

The level of signals used in this implementation is such that the term ``perfect copy"  does not apply. Whereas classical signals may admit the concept of a perfect copy (apart from technical noises), {\em any} ``copy" of a signal in the mesoscopic range produces a distinct output due to the inherent noise in the channel. In other words, use of signals where the signal-to-noise ration $S/N$ is very high ($=$``classical" signals) produces undistinguishable copies. Of course, an XOR of classical signals produces well-defined signals and the attacker's task will be solely cryptanalysis of perfectly defined signals - a purely mathematical task.
On the other hand, when dealing with an intrinsically noisy channel, the first task of the adversary task is to make sense of the signals being transmitted. Moreover, when these signals are distributed among different physical states, the adversary task to obtain even the sequence of signals to a posterior cryptanalysis is shown impossible.  Access to the $\alpha \eta$ channel gives the adversary very little information on the signals.
The amount of information available for cryptanalysis in the two cases is vastly different.

\section{The transmission and reception system}
\label{transmission and reception}

Alice has a physical random bit generator (PhRBG) that produces true random bits continuously.
One of the basic questions to be answered in this paper is: {\em Given that A and B start sharing a secret random sequence of length $c_0$,  what is secure length of bits to be extracted in the successive rounds of this system after $c_0$ have been used}?

 The original transmission protocol described in \cite{barbosa1} and \cite{barbosa2} indicate the need for a PA protocol to be applied at the transmitter and receiver stations to eliminate any information that the attacker could have obtained in these attempts but no specific procedure was proposed. Furthermore, in the present paper, instead of assuming some repetition of bases as done in \cite{barbosa1}, the idea of a ``{\em bit pool}" is used leading to a substantial improvement on the overall security of the system. This will be explained ahead.

 Basically the idea is use to PA to reduce the amount of information accessed by E to a negligible amount while giving A and B access to a refreshed bit sequence to be used.
 This frustrates even a-posteriori attacks using known-plaintext by an adversary trying to recover past bit sequences based on the bit sequences obtained from the plaintext used.
 Before explaining the protocol within the bit-pool, a description will be given of the physical system where the protocols will operate.

\begin{figure}
\centerline{\scalebox{0.5}{\includegraphics{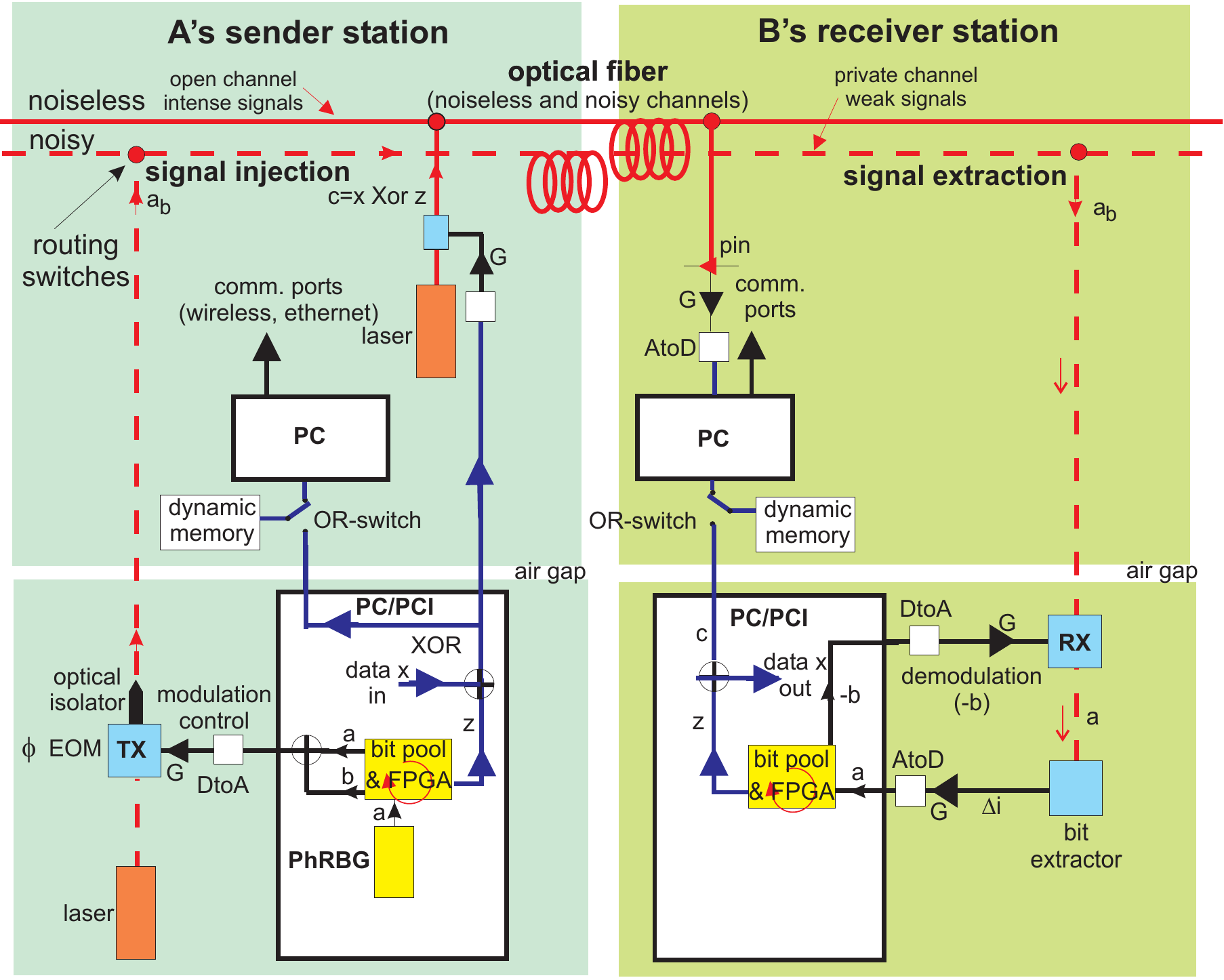}}}
\caption{ \label{AliceBobEveAirGap_23June2015} A simplified sketch of the sender and receiver stations used by the legitimate users Alice and B. The fiber channel may be a single fiber with a noiseless channel and a noisy channel. The noisy channel is used to distribute secret random bits between Alice and Bob while the noiseless  channel is used to transmit encrypted  information. These channels have to be spectrally separated to avoid spill-overs from the intense noiseless channel to the noisy channel.
The distributed secret bits are treated and privacy amplified in a bit pool in Alice's and Bob's stations. The distilled secure sequence of bits, $z$, is used for encryption of text, image or voice.  Signals are sent just once and never repeated. Actually, sender and receiver systems may be contained in both A and B stations to simultaneously offer sender/receiver capabilities.}
\end{figure}

Fig.~\ref{AliceBobEveAirGap_23June2015} sketches the main parts of the key distribution system. A single optical fiber contains an optical noisy channel ($\alpha\eta$) and a noiseless channel, {\em both} fully accessible to Alice, Bob and Eve. Signals from the laser (carrier) are modulated at station A and demodulated at station B. A physical random bit generator (PhRBG) \cite{enigma} feeds a control station composed of a computer and electronics  to perform required functions such as digital to analog (DtoA) conversion, analog to digital (AtoD), XOR, and PA operations on a
 bit pool. A final stream $z$ of secure bits is extracted from the pool to encrypt bit-by-bit any desired data $x$ (``message'') ($c=x \oplus z$) to be sent from A to B or from B to A by the public channel.
 The PhRBG continuously generates a fast stream of random bits $a$ that feeds the control station.
 \renewcommand{\dbltopfraction}{1.0}

\begin{figure}[h]
\centerline{\scalebox{0.25}{\includegraphics{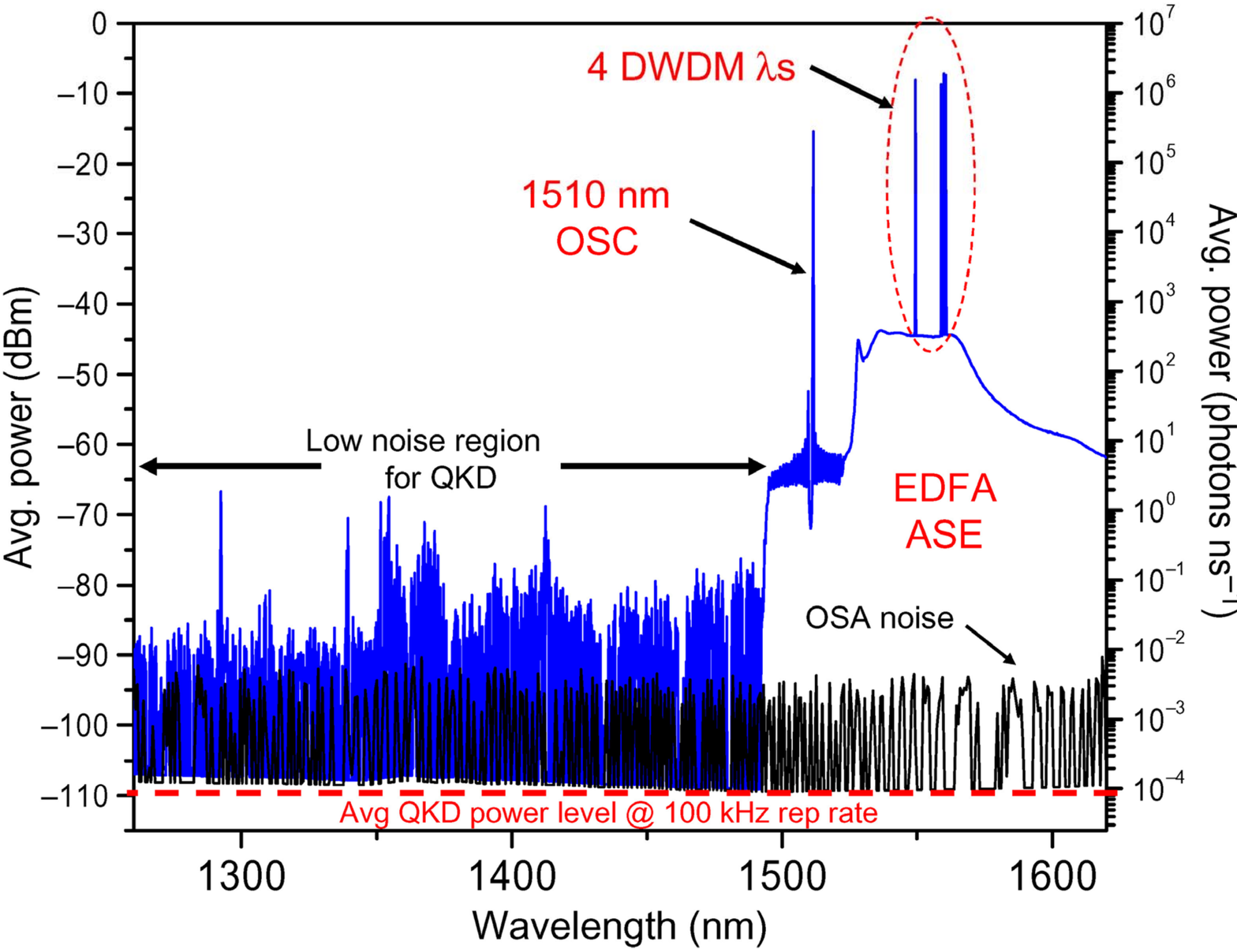}}}
\caption{ \label{SignalsAndNoises_TelcordiaAndOthers}  Different signal and noise levels due to different processes in a network.  The minimum noise level is the OSA noise due to the spectrum analyzer being used while the maximum signal is due to amplified signal that also produces a relatively high amplified spontaneous emission signal (EDFA ASE) that acts as a noise for some communication channels. Signal from an Optical Supervisory Channel (OSC) is also shown.}
\end{figure}

 Briefly described, A and B secretely shared $c_0=m \: n_0$ bits to create $n_0$ modulation bases to encode $n_0$ fresh bits generated by the PhRBG (to be discussed ahead). $m$ is the number of bits necessary to specify each basis.
Eve has full access to both channels, noisy and noiseless. The optical signals are created by phase modulation of a laser beam to create the information signals transmitted by the optical fiber.
 Signals in the classical channel  have a high signal to noise ratio (SNR), while the noisy channel has a relatively small SNR. The noisy channel is wavelength separated from the classical channel such that the wavelength separation avoids overlap of the wings of the strong signals in the classical channel with the weak signal carried by the noisy channel. Fig.~\ref{SignalsAndNoises_TelcordiaAndOthers} (taken from \cite{Telcordia}) exemplifies signal and noise levels taken at some point in a fiber network that include optical amplifiers. The mesoscopic signals used in the key distribution system being discussed are above the the QKD level (single photon level), but well below an intense (=classical) signal. We define a classical signal as a signal that can be perfectly copied or only subjected to technical noises, that could be eliminated by improved techniques.\\
Wavelength Division Multiplexing (WDM, dense or coarse) technology can be used to set distinct channels in the same single fiber around $1553$nm. WDM wavelengths are standardized with 100GHz spacing in optical frequencies, with a reference fixed at 1552.52nm (193.10 THz). DWDM can use 50 GHz channel spacing or even 25GHz spacing for up to 160 channel operation. For a small size network, where no optical amplifier is needed and only an Optical Supervisory Channel is used, the amount of noise is much less, simplifying the setting of the wavelengths to avoid cross-talk with the noisy channel carrying the bits for the key distribution protocols.

The communication protocol is presented in the next section and the operations performed by the control station will be discussed ahead. It is assumed that bits can be send in runs of size $n(i)$, where $i$ is the run index.
 \begin{figure}[h]
\centerline{\scalebox{0.2}{\includegraphics{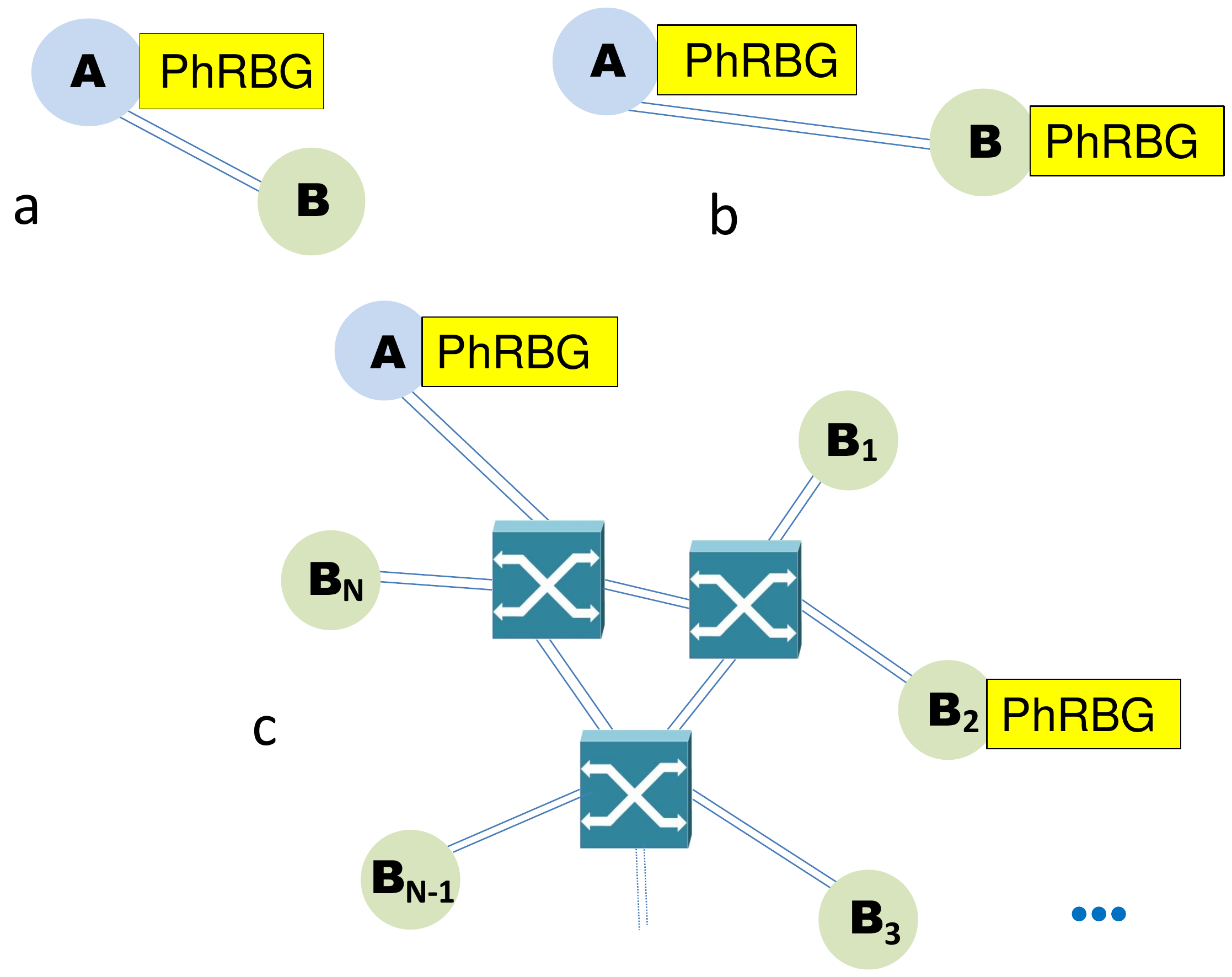}}}
\caption{ \label{Networking_3} Some networking possibilities for the key distribution platform. a) is a configuration where only A possess a PhRBG whereas in b) both A and B have equal capabilities. In c), several receiving stations can be set for a single key-sender. }
\end{figure}
It is important to realize that the separate sender and receiver station capabilities shown in Fig.~\ref{AliceBobEveAirGap_23June2015} and Fig.~\ref{Networking_3}-(a) could also be set in a same station, where both A or B have emission and receiver capabilities Fig.~\ref{Networking_3}-(b). In this case, there is no change in the logical procedures used, all arguments and explanations remain valid. Networking with more stations is also possible; see Fig.~\ref{Networking_3}-(c).
 \begin{figure}[h]
\centerline{\scalebox{0.5}{\includegraphics{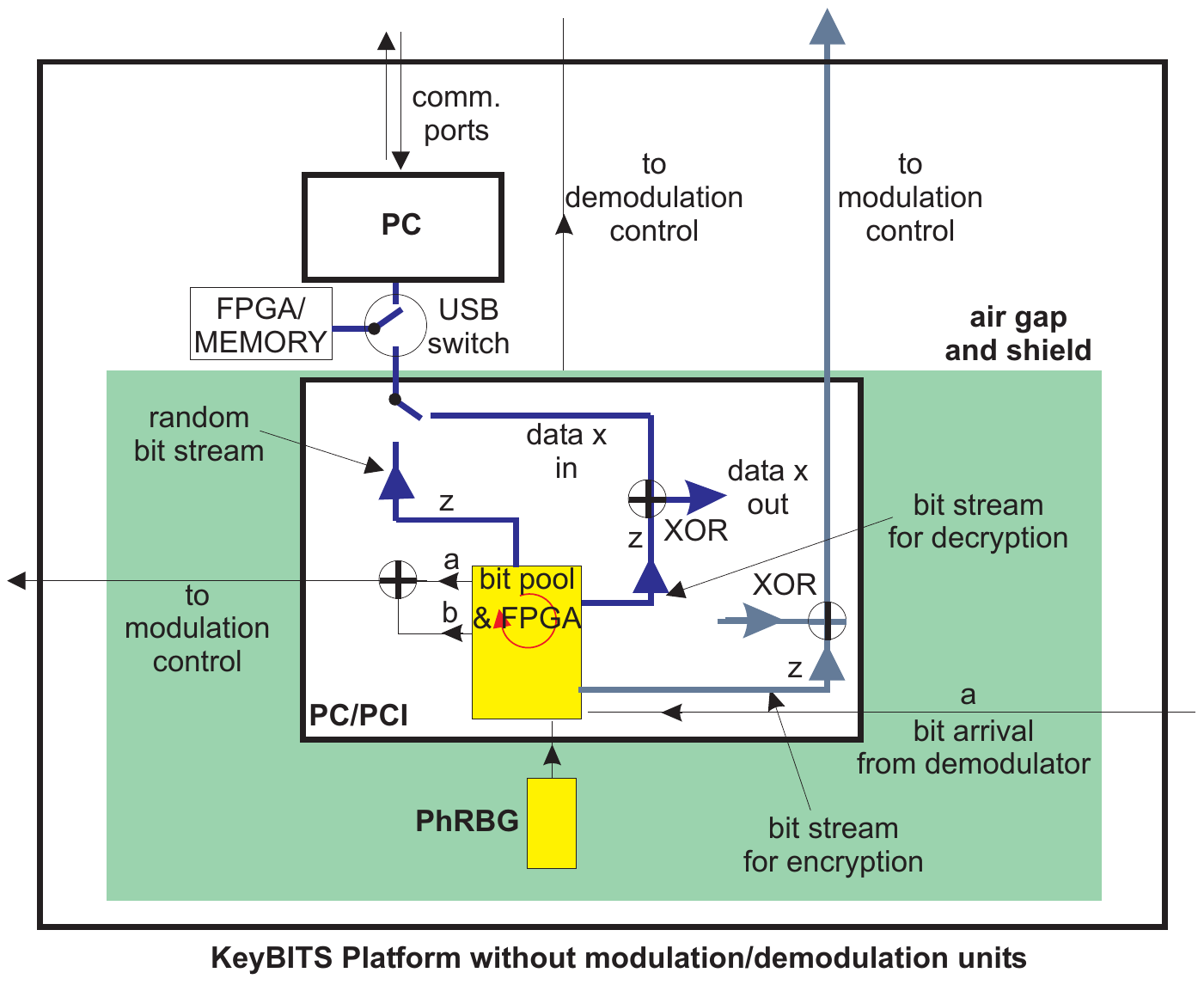}}}
\caption{ \label{KeyBITSPlatform_AirGap_21Jan2014} KeyBITS modular secure communications platform. See Fig.~\ref{AliceBobEveAirGap_23June2015} for more details.}
\end{figure}

In Fig.~\ref{AliceBobEveAirGap_23June2015} a dashed line indicates an ``air-gap" that separates a  region (bottom part) assumed free of undesired interferences of any kind, including reception or emission of electromagnetic and acoustical signals. All information going to the secluded region by the air-gap has to be monitored so that only electronic signals with a fixed format and {\em authentication} are allowed. This monitoring may happen through a FPGA (Field-Programmable Gate Array) and memory accessed  by one of the two computers, in {\em exclusive} operations, through an automatized USB (Universal Serial Bus)-switch. When operated with the computer in the air-gap, the FPGA tag the files using the authentication-hash adopted before sending them to the other computer.

In fact, separated emitter and reception stations in Fig.~\ref{AliceBobEveAirGap_23June2015} may be together in both stations A and B, so that each one is autonomous, in a modular unit. Fig.~\ref{KeyBITSPlatform_AirGap_21Jan2014}  shows a modulus of this platform, without the modulation and demodulation units.

\section{The physical protocol}
\label{protocol}
As shown in \cite{barbosa1} and \cite{barbosa2}, and further discussed in Section~\ref{transmission and reception},  the noise in the channel combined with the use of closely separated bases
reduce enormously the probability of success of Eve. The signals she obtains do not allow her to obtain reliable bit sequences to be analyzed. This is the physical protection level in the key distribution scheme. However, this is not the only level of difficult existing in the system.

A first round of sending bits will be described to establish the basic ideas.
A PhRBG continuously generates random bits $a$ that can be processed in a bit pool with operations fastly processed by a  FPGA or ASIC (Application-Specific Integrated Circuit).

Initially, this bit pool starts with the shared $c_{0}$ random bits, constituted of $c_{0}=m \: n_0$ bits to create $n$ modulation bases.  For the sender, the total number of bits in the beginning of the process is then $n_0+n_0 m$. It should be emphasized that despite the need of $m$ bits to create a basis for modulate {\em one} bit, the process has been demonstrated to be very fast in hardware.

The choice of $m$ depends on the physical choice of the bases to be used and the intensity of light, or average photon number $\langle n \rangle$ in the noisy optical channel (see \cite{barbosa1} for explanations). For example, if optical {\em phase} values are used in a circle of 2$\pi$ values, a choice of $M$ values implies a distance of $\pi/M$ between bases. A bit 1 could be represented by a phase value $\phi_1=\pi$ while a bit 0 is given $\phi_0=0$ in one given basis $n_M$. In the other closest bases $n_M\pm 1$, the opposite choice is adopted, alternating ones and zeros.

The idea \cite{barbosa1} is to set the light's noise such that it overlaps several physical  bases. The choice of the number of bases $M$ is based on a POVM --Positive Operator Valued Measure-- which defines the probability of error given to the attacker (See Section V of \cite{barbosa1}) and it will be directly connected to the average number of photons $\langle n \rangle$. Once $M$ is defined, physical signals are generated creating a  modulation (say, a phase $\phi$) upon the laser beam. This physical modulation is being called an encryption basis for a fresh bit. To create {\em each} basis,
 $m$ bits are necessary, $2^m=M$ or $m=\log_2 M$:
\bea
&&\!\!\!\!\!\!\!\!\mbox{phase basis number}(\mbox{for}\:\:\phi_b = 0 \frac{ \pi}{M} ,1 \frac{ \pi}{M},\cdots (M-1)\frac{ \pi}{M})\nonumber\\
                           &\equiv & b(m) 2^{m-1}+b(m-1)2^{m-2}+\cdots b(1)2^0\:.
\eea
All bits of $a$ generated by the PhRBG are represented by phase values $\phi_a$ (either $\phi_1$ or $\phi_0$) and added to the corresponding phase $\phi_b$ associated with the basis being used:
$\phi_a+ \phi_b$ and sent from A to B.

With $n \: m$ bits, $n$ modulation bases $b$ are created. They modulate the $n$ fresh bits of $a_0$: $a_0 \oplus b$. As this sequence $b$ is known to A and B, B could used it to demodulate the received sequence, extracting $a_0=b\oplus(a_0\oplus b)$.

As a brief comment, in the BB84 protocol, two bases are defined to send one bit that is carried by a single photon. The adversary must not know in which of the two bases the bit was encoded. In a parallel way, for the key distribution protocol discussed in this paper, the optical noise must protect against attempts by the adversary to know which basis was used.

Now, A and B share the sequence $a_0$. Eve may have obtained some {\em statistical} information $t$ on these bits and A and B task is to eliminate $t$  by PA -- therefore, calculation of $t$ is essential.
This is shown ahead.

Another level of difficulty, computational, will be added -- usually, this level of difficulty is used as a stand-alone protection level and may be sufficient by most of the cases even with noiseless signals. This mathematical level of protection will be discussed ahead as well as the effect of combining these two protection levels.

{\bf Physically leaked bits --}
Assume that Alice sends $n$ ($n=\mbox{Length}[n_0]$) uniform random bits to B. Eve has complete access to the transmission channel, close to the sender, where the signal is maximum, and disposes of the ideal equipment, subjected to the laws of Physics, to measure and record  all emitted signals. No one monitors Eve's intrusion and will not constrain her endeavor in anyway. However, Alice will not send bits in a repeated way; every bit information is sent {\em only once}. This way, Eve obtains noisy signals representing the sent bits and will treat them individually or collectively, as she pleases.

As shown by the POVM calculation in Section V of \cite{barbosa1}, there is a minimum probability of error $P_e$ for Eve when measuring any bit due to the inherent noise in the optical channel and the $M$-ry bases used. $P_e$ is a function of the average photon number  $\langle n \rangle$ in the signal representing a bit  and  $M$, the number of bases used in the $M$-ry communication protocol. The POVM calculation utilizes the wavefunction or density matrix representing {\em all} information about the transmitted bit. This is the maximum amount of information available about a physical system. The result of the calculation indicates the best Eve could obtain, even with an ideal measuring system and analyzing capabilities.

For numerical examples of these results, see Fig. 3 in \cite{barbosa1}.
The probability of having a correct bit assignment by Eve is $P_r=1-P_e$. Therefore, Eve is able to {\em statistically} assign correctly or ``extract'' $t\equiv t_{\mbox{bit}}=P_r-0.5=0.5-P_e$ of each bit. \\Therefore, $t_{\mbox{bit}}$ is a extraction rate of Eve (or, leak per bit).

\begin{figure}[h!]
\centerline{\scalebox{0.25}{\includegraphics{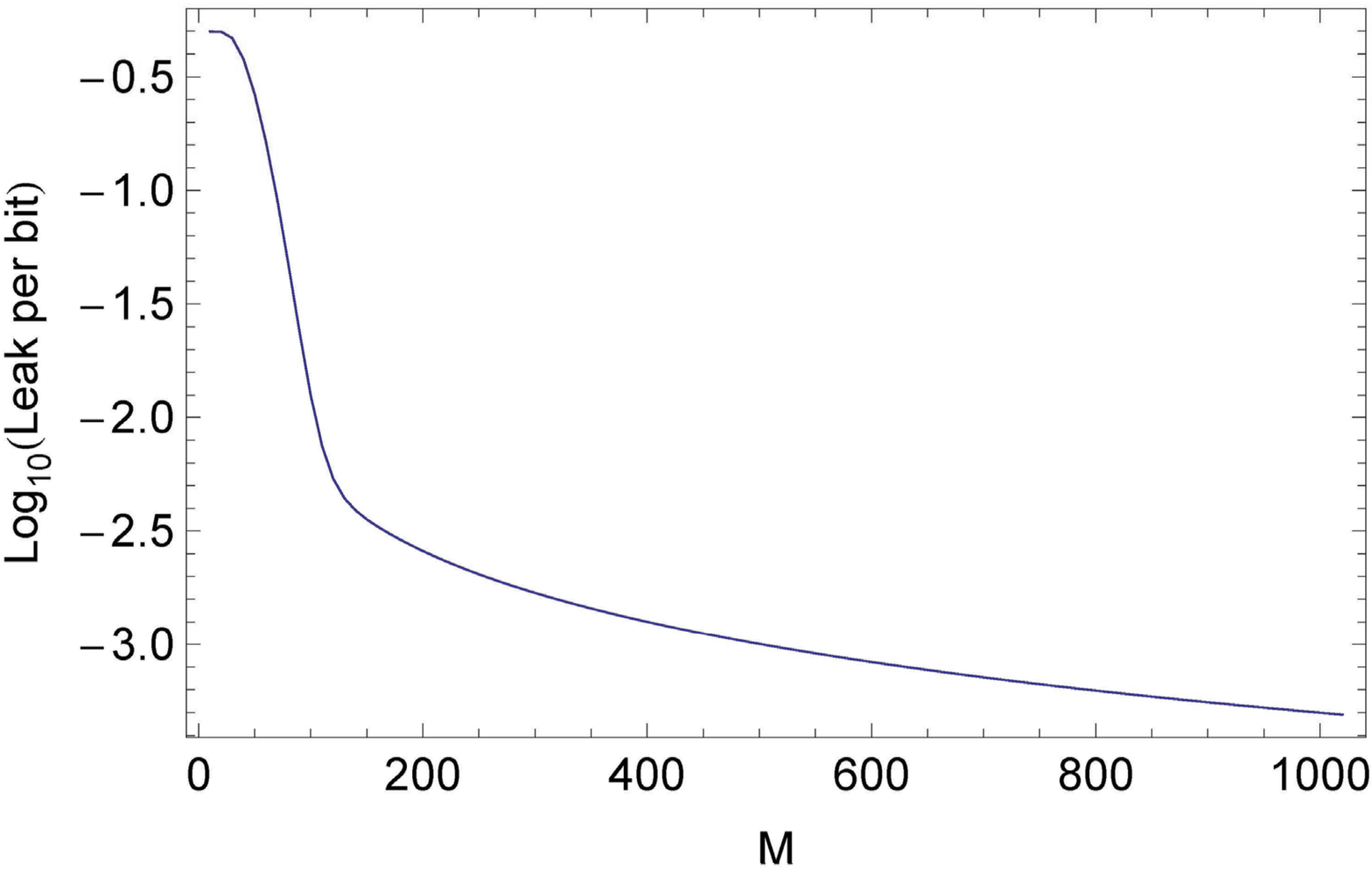}}}
\caption{ \label{LogpDelta}
Logarithm of the loss per bit to Eve, $t_{\mbox{bit}}$, as given by the POVM calculation that provides $P_e$. Here $\langle n \rangle=1000$.
 }
\end{figure}

Fig.~\ref{LogpDelta} exemplifies the behavior of $\log_{10}t_{\mbox{bit}}$ as a function of $M$.
As will be shown in Section~\ref{PA}, Eve´s probability to obtain all bits in a sequence is completely negligible.

\section{Signal Modulation and Demodulation}

The left side of Fig.~\ref{AliceBobEveAirGap_23June2015} shows a modulation system (at Alice's station) that injects signals in the ``noisy" channel of the optical fiber. At the right side of the same Fig. (at Bob's station) a demodulation system extracts the signals sent by Alice, erasing the signals representing the encoding bases that produce the indistinguishability of the signals to the attacker.

The modulation and demodulations systems are discussed in \cite{ModulationDemodulation} and will not be discussed here. The final signal $\Delta i=i_e-i_f$ representing the bits as extracted by Bob, come from the two pin detectors in the demodulation system and are proportional to the streams of photons
\bea
\!\!\!\!\!\langle n_e \rangle \!\!\!&=&\!\!\!-\frac{1}{4} |\alpha|^2 \!\!\left[\!\sqrt{3} \sin \varphi  \cos
   ^2\left(\frac{\Delta}{2}\right)\!\!+\!\!\sqrt{3} \cos \varphi  \sin \Delta\!-\!2\right]\\
\!\!\!\!\!  \langle n_f \rangle \!&=&\! \frac{1}{4} |\alpha|^2 \!\!\left[\!\sqrt{3} \sin \varphi  \cos
   ^2\left(\frac{\Delta}{2}\right)\!\!+\!\!\sqrt{3} \cos \varphi  \sin \Delta\!+\!2\right],
\eea
where $\Delta$ is the path phase difference between the two arms of a fiber Michelson interferometer in the bit extractor (see Fig.~\ref{AliceBobEveAirGap_23June2015}).

\begin{figure}[h!]
\label{current}
\centerline{\scalebox{0.3}{\includegraphics{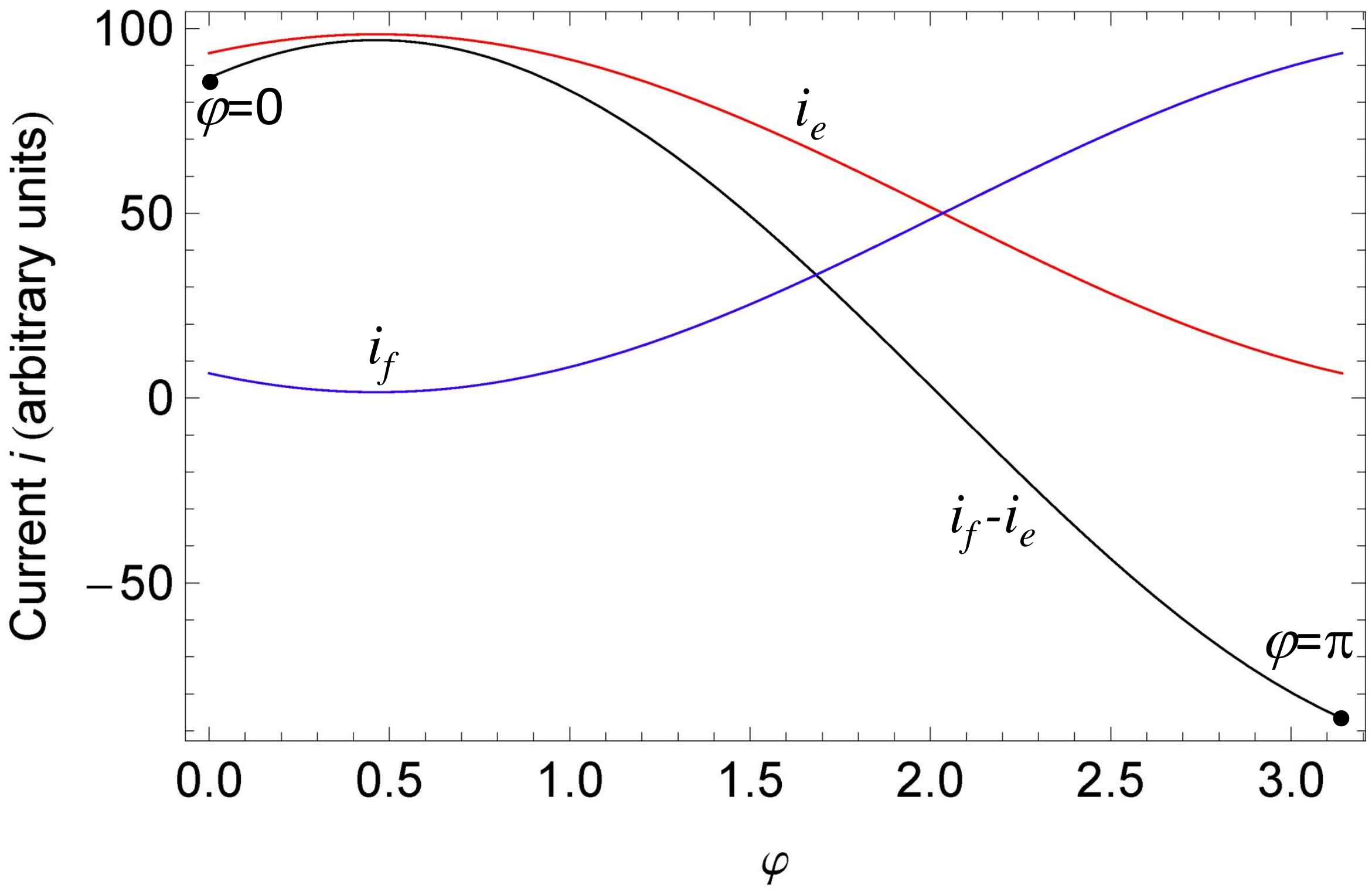}}}
\caption{ \label{MichelsonCurrents} Photon currents $i_e$ and $i_f$ obtained from the pin diodes and the final difference current $\Delta i=i_f-i_e$ as a function of the input phase $\phi$. Particular values are indicated with $\phi=0$ and $\phi=\pi$ that represent bits 0 and 1.}
\end{figure}

Fig.~\ref{MichelsonCurrents} shows plots for the direct currents  $i_e$ and $i_f$  for a given laser intensity (arbitrarily taken at $|\alpha|=10$, and  $G=1$, $\eta_d=1$ and a unitary time interval. $\Delta$, set by the piezoelectric driver, is set at $\Delta=\pi/2$); $G$ is the detector's electronic {\em gain} and $\eta_d$ is the detector's {\em efficiency} in the photon-to-electron conversion.
It is seen that the best resolution for bits 0 and 1 is obtained from the difference of the two currents, or $\Delta i=i_f-i_e$ and not from either current outputs $i_e$ or $i_f$ alone.

As was shown, physical noise can create a physical barrier to the attacker making it impossible for him/her to extract clean bit  signals from the channel. The attacker degree of success is almost negligible and, furthermore, this small probability of success is {\em statistical} in character. \\
At the same time the legitimate users extract clean signals and obtain the bits sent from the current at the demodulator, as given by Fig.~\ref{MichelsonCurrents}.

These different results for the attacker and for the legitimate user can be understood by the overlap between two states, as given by Eq.~(\ref{overlap}). For the legitimate user he has only to distinguish between a bit $0$ and a bit $1$ encoded by a known basis value $b$, within $M$ possible values. For him, $\Delta \phi=\pi$ and, therefore
\bea
\langle \Psi(\phi)|\Psi(\phi+\pi) \rangle=e^{-2 |\alpha|^2} \rightarrow 0\:.
\eea
This result shows a comparison between two almost orthogonal states, of easy identification, with negligible overlap. On the contrary, not knowing the basis value the adversary has a complex measurement problem that limits his/her knowledge according to what was shown by the POVM calculation in \cite{barbosa1}.

\section{Increasing the protection level}
\label{increasing}

The security of the proposed key distribution system does not stop at this physical barrier but has its security further strengthened by Privacy Amplification (PA).
The final security then rests on a combination of physical and mathematical protections. Each of these aspects can be calculated as well as the final probability for the attacker to obtain the final bit sequences being shared by A and B. This provides a rigorous proof of the security of the system. The adopted PA protocol will be discussed in the next sections.

Besides the security features for signal protection, the system also uses a Message Authentication Code (MAC) to  guarantee that  a message sent to Bob  by Alice was really sent by Alice and not by someone else.

An OMAC procedure, a  MAC that uses just {\em one} key $K$ and with high security is adopted \cite{IwataKurosawa}.  The information to be authenticated is divided in $M(j)$ blocks and the key $K$ is applied to the first block. The output, together with the key $K$, is applied to the second block and so on. At the end of the operations a tag ``T'' is produced. The final output, with the tag, is send to the receiver. He applies the same procedure and generates  a tag. If the generated tag coincides with $T$, the information is authentic. There is no need to decrypt the message to perform the tag check.

\section{Privacy Amplification protocol}
\label{PA}

Before discussing the PA protocol, is should be emphasized that although the physical protocol uses a somewhat larger number of bits ($\log_2 M$) to encode {\em one} bit, the process is continuously sustained in rounds of $s$ bits, in an unlimited way. This process has been shown very fast in hardware.
\begin{center}
{
\begin{table}
\caption{Privacy Amplification protocol for the KeyBITS platform}
\label{PAtable}
\begin{tabular}{|l|c|c|}
  \hline
  \multicolumn{3}{|c|}{\small \bf PA protocol}\\
  \multicolumn{3}{|l|}{\small{\bf INITIALIZATION:} A and B share $c_0$ of size and entropy $m\: s$.}\\
  \hline
  \hline
  \multicolumn{3}{|l|}{\small\bf ALICE}\\
  \hline
   {\small $\#$} &{\small ACTION}&{\small COMMENT } \\
   \hline
  {\small 1a} &{\small  $a_i=\mbox{GetString(PhRBG)}$ }&{\small get bitstring from PhRBG}\\
  {\small 1b} &{\small  $b_i=c_{i-1}[1,m s ]$ }&{\small extract $m \: s$ from pool for bases $b$}\\
  {\small 1c }& {\small Code\&Send$(a_i,b_i)$ }&{\small send over $\alpha\eta$ channel}\\
  \hline
  {\small 2} &{\small  SendCC$(f)$  }         &{\small send instance of universal hash $f$}\\
  &&{\small over classical channel}\\
  \hline
  {\small 3a} &{\small $ c_i=f(c_{i-1}||a_i)$ }&{\small Alice applies PA from $ms+s$ bits }\\
  &  &{\small to $m s+s-t-\lambda$}\\
  {\small 3b} & {\small $c_i=f(c_{i-1}||a_i)$ }&{\small Alice uses $\overline{s}=s-t-\lambda$}\\
                 &            &{\small bits from pool as the key stream $z$.}\\
                  &           &{\small The remaining $m\: s$ bits form }\\
                   &          &{\small the bases' bits for next round.}\\
  \hline
  \hline
  \multicolumn{3}{|l|}{{\small \bf BOB}}\\
  \hline
 {\small 1a} &  &{\small no matching step to Alice's}\\
{\small  1b} & {\small $ b_i=c_{i-1}[1,m s]$}&{\small get bases bits from initial pool value} \\
 {\small 1c} &{\small  $a_i=$Receive\&Decode$(b_i)$}&{\small receive bits from $\alpha\eta$ channel} \\
 \hline
 {\small  2} & {\small  ReceiveCC$(f)$ }&{\small receive instance of universal hash $f$}\\
  \hline
{\small  3a} & {\small $c_i=f(c_{i-1}||a_i)$ }&{\small Bob applies PA from $ms+s$ bits} \\
  &  &{\small to $ms+s-t-\lambda$}\\
{\small  3b} &{\small  $z_i=$}&{\small Bob uses $\overline{s}=s-t-\lambda$}\\
      &        {\small $\!\!\!\!c_i[ms+\!\!1,ms\!+s\!-\!t\!-\!\lambda]\!\!\!\!$}               &{\small bits from pool as the key stream $z$.}\\
                  &           &{\small The remaining $m\: s$ bits form }\\
                   &          &{\small the bases' bits for next round.}\\
  \hline
\end{tabular}
\end{table}
}
\end{center}

The Privacy Amplification protocol adopted  uses a bit pool of constantly renewed random bits. For details, see \cite{jeroen}.  Fig.~\ref{AliceBobEveAirGap_23June2015} can be used as a reference for description of the PA protocol.
Before discussing the  level of security, a summary of the PA protocol steps is given in
Table~\ref{PAtable}. After this summary, conditions for its applicability will be discussed.

The first round of the protocol will be described in words:\\
INITIALIZATION:
Alice and Bob share a starting sequence $c_0$ of secret random bits. The sequence has size $c_0=m\: s$, where $m$ is the number of bits necessary to describe one of the $M$ basis and $s$ is the size of the first fresh sequence of random bits to be shared between A and B.

{\bf Alice first steps --} \\
A1a: Alice gets a random bitstring of length $s$ in the bit pool fed from the PhRBG.\\
A1b: Alice gets the shared starting sequence $c_0$ and partition it in $s$ parts with $m$ bits each. Each subsequence of length $m$ randomly specifies one basis among the $M$ bases.\\
A1c: Alice encodes each bit in $s$ with the corresponding basis and sends the signal to Bob over the noisy channel. See Section \ref{protocol} for a description of the physical modulation to be used.
  Beforehand, Alice and Bob had agreed on the PA's security parameter $\lambda$  and calculated the statistical fraction $t$ of a bit ($t\equiv t_{\mbox{bit}}$) leaked to Eve, see Fig.~\ref{LogpDelta}  and Eqs.~(7) to (12) in Ref.\cite{barbosa1}.\\
A2: Alice sends an instance of a universal hash function $f (\in \: \mathcal{F})$ to Bob over a noiseless channel with public acess.\\
A3a: Using $f$ Alice applies PA and reduce the total number of bits $s+m s$ to $\overline{s}=s+m s-t-\lambda$.
The eavesdropper has {\em no} knowledge on $\overline{s}$ or on the modified $m s$ sequence. See \cite{jeroen} for details. \\
A3b: The reduced sequence $\overline{s}$ is the distilled fresh random sequence $z$ to be used for OTP encryption. The remaining fresh random sequence $m\: s$ will form the bases for the next run.\\
{\bf Bob first steps --} \\
B1a: There is no corresponding step to Alice's 1a.\\
B1b: Bob gets the shared starting sequence $c_0$ and partitions it in $s$ parts with $m$ bits each. Each subsequence of length $m$ randomly specifies one basis among the $M$ bases. Bob and Alice are then using the same set of bases.\\
B1c: Bob receives the physical signals sent by Alice, demodulates them (See Fig.~\ref{AliceBobEveAirGap_23June2015}) and obtain the random bit stream coded with the $s m$ bases. As Bob knows the bases coding, he decodes the random stream and obtains the stream $s$ sent by Alice.\\
B2: Bob receives $f$ over the classical channel.\\
B3a: Using $f$ Bob applies PA and reduce the total number of bits $s+m s$ to $\overline{s}=s+m s-t-\lambda$.\\
B3b: The reduced sequence $\overline{s}$ is the distilled fresh random sequence $z$ to be used for OTP encryption. The remaining fresh random sequence $m\times s$ will form the bases for the next run. Therefore, both Bob and Alice share a secure sequence of random bits $\overline{s}$ to be used as OTP.

This means that the generated stream from the PhRBG at Alice's station was transferred to Bob and a distilled secure sequence of random bits and base bits is obtained.\\
The protocols proceeds to next similar runs. After $n$ runs, Alice and Bob share $n \overline{s}$ bits.

{\bf Preliminary conditions for the PA protocol --} The  protocol for Privacy Amplification \cite{BBCM} (or PA) offers a powerful tool to decrease the amount of information an adversary (E) might have acquired on a bit string transmitted from one legitimate user (A) to a second one (B).

In this paper, A sends $n$ random bits to B, from which E is able to statistically gain $t_{\mbox{bit}}$ of information per pulse sent. The amount of gained information by the adversary over a string of $n$ randomly distributed bits
is $t_{\mbox{bit}}^{n}$.

Among quantities or conditions that the PA protocol need to be applicable are the statistical gain $t_{\mbox{bit}}$ as well as a bound on the
second-order conditional R\'enyi entropy, $R_2(n|V=v)$, as seen by the adversary E. Here $V$ designate the variable under control of E that has some degree of correlation to $n$.

Some preliminary steps may help to calculate $R_2(n|V=v)$. The ``collision probability" for a variable $X$ specified by a probability distribution$P_X(x)$ can be defined (see \cite{BBCM} for details) as
\bea
P_c(X)=\sum_{x {\Large \epsilon} \chi}P_X(x)^2 \:\:,
\eea
from which the R\'enyi entropy of $X$ can be calculated:
\bea
R(X)=-\log_2 P_c(X) \:.
\eea
The {\em entropy} in binary digits is also given in ``bit" units, and may be fractionary, differently from the physical bits 0 or 1 (encoded physical signals). Given an event $\mathcal{E}$ on $X$, with conditional probability
$P_{X|\mathcal{E}}$, one may directly write the collision probability
$P_c(X|\mathcal{E})$ and the conditional R\'enyi entropy
\bea
R(X|\mathcal{E})=-\log_2 P_{X|\mathcal{E}}^2 \:.
\eea
The variable $X$ of interest should represent the bit stream transmitted from A to B and the event $\mathcal{E}$ represents Eve´s access to that stream.

\begin{figure}[h!]
\label{current}
\centerline{\scalebox{0.25}{\includegraphics{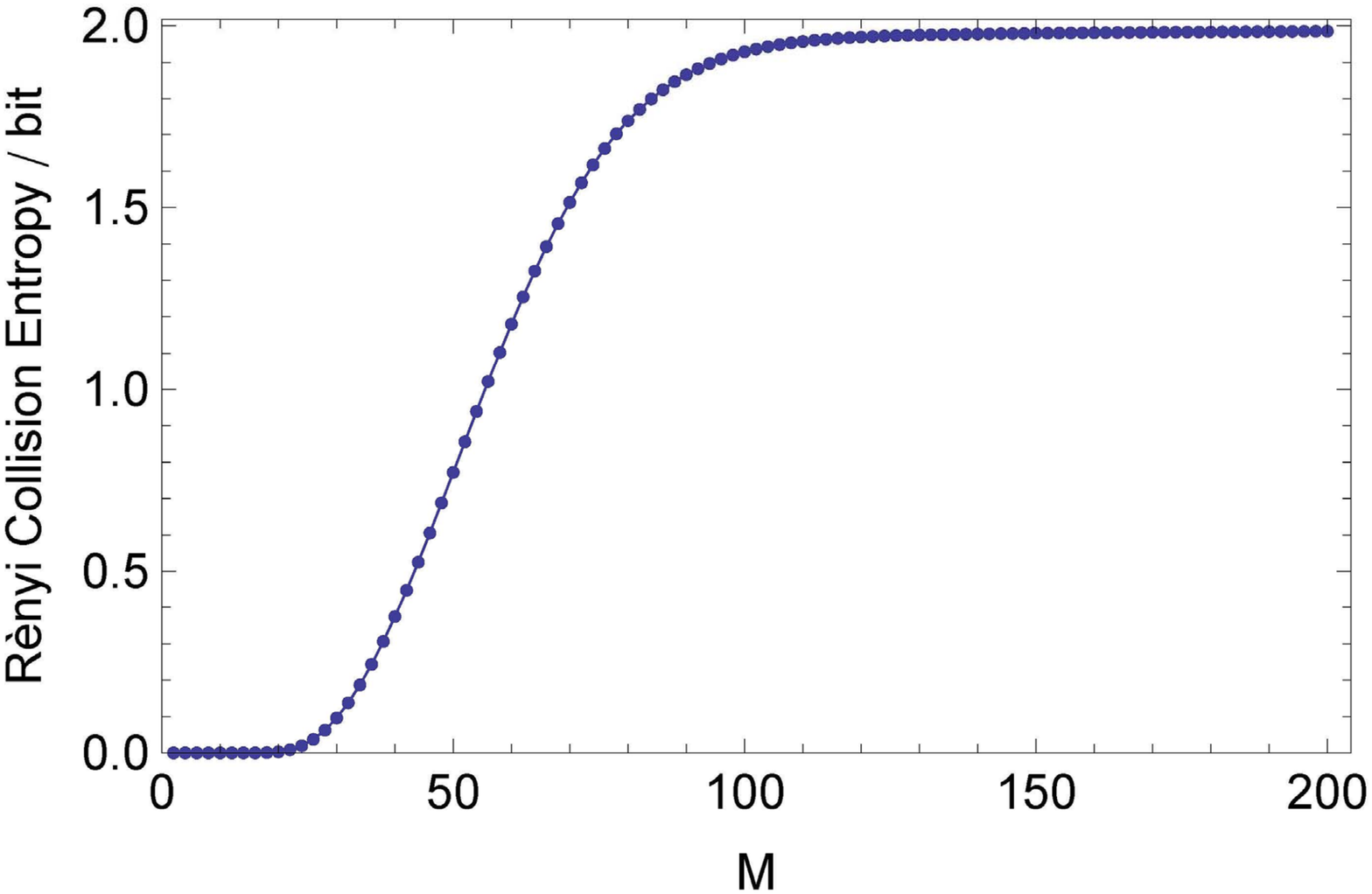}}}
\caption{ \label{pH2} Example of the Conditional R\'enyi collision entropy of order two for bits being transmitted with $\langle n \rangle=100/$bit signal as a function of the number of bases $M$. }
\end{figure}

The mapping of $P_{X|\mathcal{E}}$ in terms of the physical processes
 gives
\bea
P_{X|\mathcal{E}} \rightarrow P_r=1-P_e \:,
\eea
and
\bea
\label{Rx}
R(X|\mathcal{E})=-\log_2 P_{X|\mathcal{E}}^2=-\log_2 \left( 1-P_e \right)^2
\eea
Fig.~\ref{pH2} shows an example of the collision R\'enyi entropy given by Eq.~(\ref{Rx}) for the case of bits being transmitted with an average number of photons $\langle n \rangle=1000$ per bit as a function of the number $M$ of bases used.
The asymptotic limit for R\'enyi entropy is
\bea
R(X|\mathcal{E})=-\log_2 \left( 1-P_e \right)^2 \rightarrow 2\:\:/\mbox{per bit}\:\:,
\eea
for $P_e \rightarrow 1/2$.
This result can be interpreted as follows: For large $M$,  Eve succeeds in obtain {\em one} collision, or statistical-success,  in every two trials.
Due to the uniformity of the random sequences, for a stream of $n$ bits the R\'enyi entropy will give the corresponding limit
$R(n|\mathcal{E})\rightarrow  2\times n \equiv c$.

The PA protocol using a compression function $G$ within a universal class of hash functions maps the received stream $\{1,0\}^n$ onto $\{1,0\}^r$. Assuming that A and B uses $z\equiv \{1,0\}^r$ as their secret stream of bits, it is known that \cite{BBCM}
\bea
H\left(z|G,V=v  \right) \geq r- \frac{2^{r-c}}{\log_e 2}\:.
\eea
Therefore, as $r < n$ and $c=2 n$, then $r < c$.  Eve's entropy on  the keys is
\bea
\label{small}
H\left(z \right)-H\left(z|G,V=v  \right)\simeq  \frac{2^{r-2 n}}{\log_e 2}\:,
\eea
and goes exponentially to zero as $n$ increases. In conclusion, the string  of $r$ random bits can be protected by the PA protocol.

\begin{figure}[h!]
\label{current}
\centerline{\scalebox{0.25}{\includegraphics{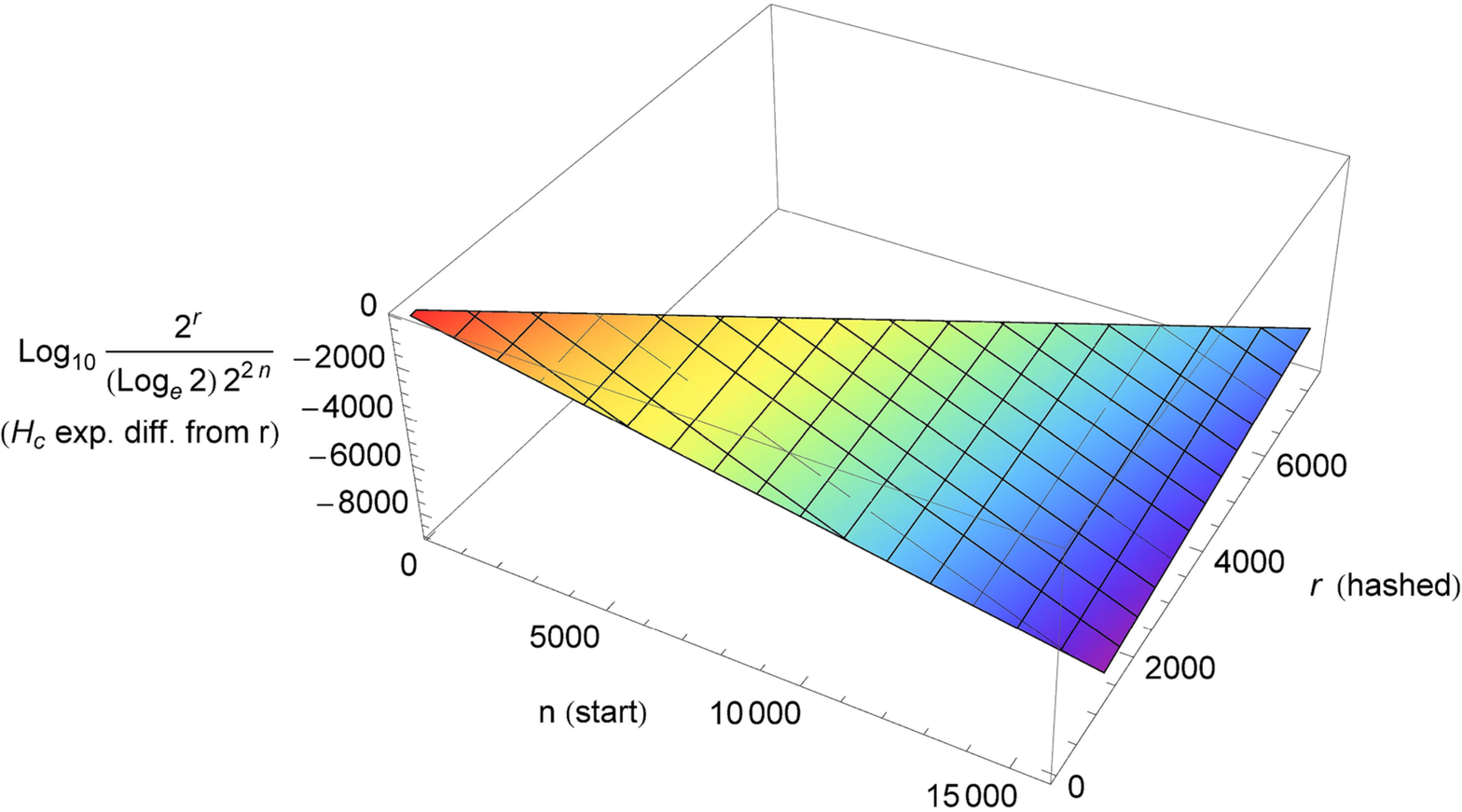}}}
\caption{ \label{pnrP}  An example that shows that Eve's amount of information on the hashed key stream is negligible (see Eq.~(\ref{small})). }
\end{figure}

Fig.~\ref{pnrP} exemplifies Eq.~(\ref{small}) for a range of $n$ and $r$ values.

\subsection{Overall security}

 Having demonstrated the possibility for application of the PA protocol \cite{BBCM} for this key distribution scheme, one may invoke
corollary 5 of the PA theorem.  In words,  the expected information of Eve about the secret key (assume length $r_t$) is given by the
mutual information $I$ on the secret key given the information $t$ acquired by Eve when Alice and Bob use a randomly chosen function from a universal class of hash functions:
\bea
I \leq \frac{1}{\ln 2 \times 2^{\lambda}}\:\:,
\eea
where $\lambda$ is a security parameter  $\lambda< r_t-t$. This way, by eliminating $t$ bits of $r_t$, Eve's information decreases exponentially while Alice and Bob knows $\overline{s}-1/(\ln 2 \times 2^{\lambda})$ bits.

In a sequence of $s$ bits  sent, two factors will work against Eve. The first one is the effect of the noisy channel on her measurements and the second one is the result of applying the Privacy Amplification protocol.

Using the probability of an error by
Eve, $P_e$, as shown in Section V of Reference \cite{barbosa1}, to correctly guess a particular bit sent through the noisy channel, the  ``hit'' probability $t_1$ is $t_1=1-P_e$. This says that Eve's probability of obtaining all $s$ bits is $t_s=\left( 1-P_e \right)^s$,
because the keys are uncorrelated as well as the physical signals that carry them. This probability gives Eve a negligible chance of success.

The legitimate users may adopt the strategy of defining the key sequence length $s$ such that after sending {\em all} of them, statistically the adversary could have gained less than one bit, that is to say $\label{short t}
s (1-P_e) < 1$.
In other words, the legitimate users choose $s< 1/(1-P_e)$. This says that the amount of $t+\lambda$ bits to be reduced from $s+ms$ (see Step 3a in Table~\ref{PAtable}) would be $t+\lambda\simeq \lambda$.

The physical noise in the channel basically reduces enormously the amount of information that Eve could obtain from the channel. On the other way, if A and B use long sequences such that $s (1-P_e) \gg 1$, a larger number of bits have to be added to $\lambda$ to make effective the PA protocol.  Therefore, from now on Eq.~(\ref{short t}) will be adopted as a condition to establish the lengths of the key sequence runs.

\begin{figure}[h!]
\centerline{\scalebox{0.25}{\includegraphics{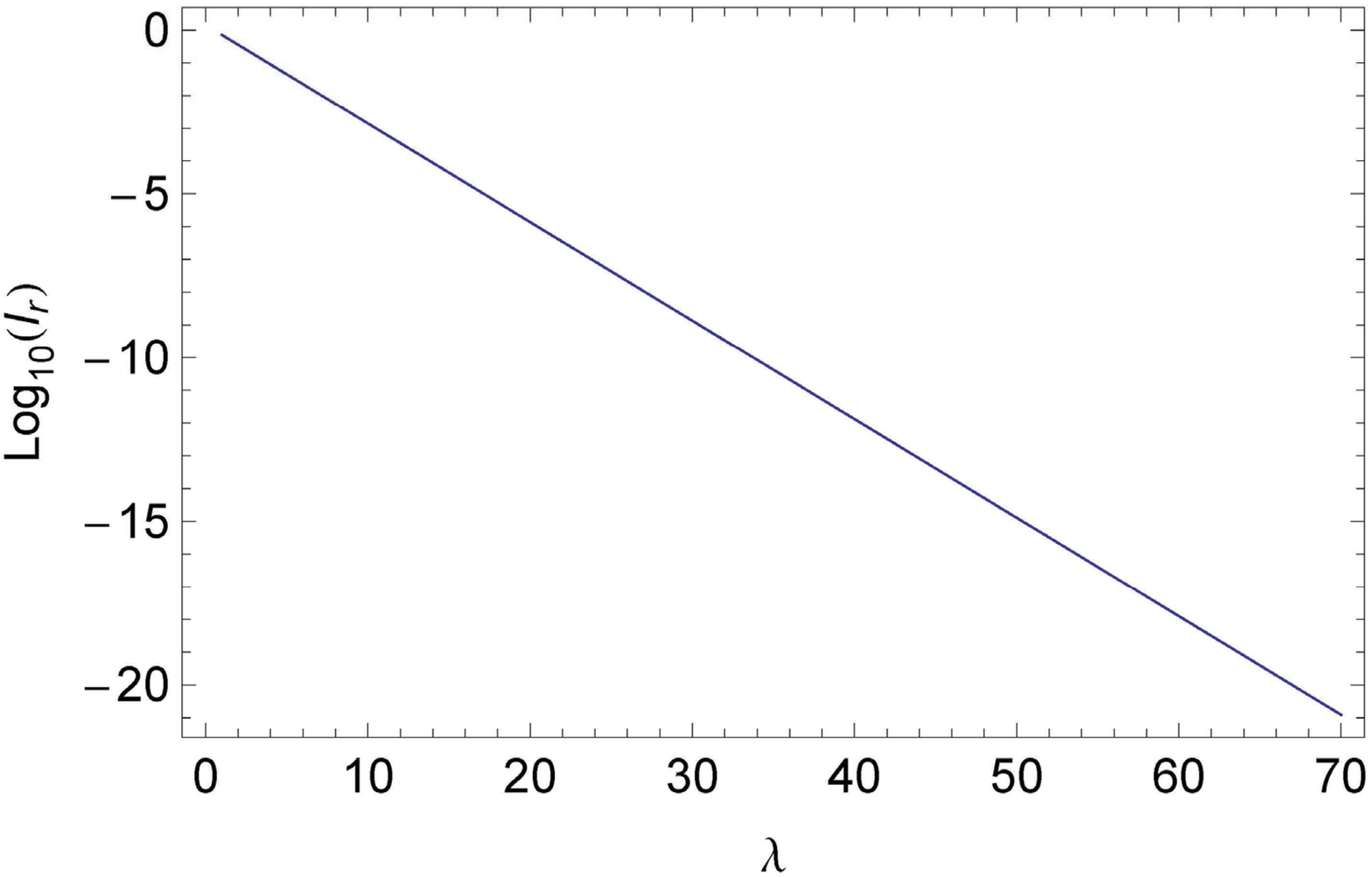}}}
\caption{ \label{pPe0} Eve´s information on $r$ bits after Alice and Bob applies the PA protocol on the bit stream  obtained from the noisy $\alpha\eta$ channel under the condition $s (1-P_e) < 1$.
 }
\end{figure}

After A and B have applied the PA protocol, together with the effects of the noisy channel, the number of bits is reduced from $s$ to $r$ distilled bits.  The information obtained  by Eve is given by
\bea
\label{s-hits}
I_{ \mbox{\tiny $r$}    }\simeq \frac{1}{\ln 2 \times 2^{\lambda}}\:\:.
\eea
Fig.~\ref{pPe0} gives an example of Eq.~(\ref{s-hits}).

\section{Adversary with information on the bit pool?}

It may be argued that even if the adversary tries to obtain the basis that has encoded a  bit and fails, some exclusive information on the bases' wheel is gained. This information may be seen as a set of bases to be {\em excluded} from the bit pool and, therefore, will simplify a posterior analysis. Therefore, the question {\em ``Will this gained information increases Eve's knowledge on the bit pool for posterior analysis on a reduced set of unknowns?"}

This question may be answered with the Mutual Information between B (or A) and E, $I(B;E)$. More specifically, assuming that A nd B utilized (secretly) a given basis $\phi_b$ to encode a bit $b$, $0$ or $\pi$, what will be the Mutual Information $I(\phi_b;\phi_E)$, where $\phi_E$ is Eve's estimated value obtained from an arbitrary measurement?

First of all, the Mutual Information will be calculated to reveal the amount of information the adversary could obtain from a bit {\em only} considering the optical noise effect on the mesoscopic signal. This absolute measure could be compared with Minimum Probability of Error by Eve $P_e^E$ already calculated in \cite{barbosa1}. As discussed, any small amount of information leaked by the channel can be privacy amplified.

\subsection{Mutual Information}
In order to write the Mutual Information
\bea
I(X;Y)=H(X)-H(X|Y)
\eea
on the desired variables, one may start with the relationships
\bea
H(X|Y)&=&\sum_{x,y}p(x|y)\log_2 \frac{1}{p(x|y)}\\
H(X)&=&\sum_{x}p(x)\log_2 \frac{1}{p(x)}\rightarrow \sum_{k=0}^{M-1} \left(\frac{1}{M}\right) \log_2 \frac{1}{\left(\frac{1}{M}\right)} \nonumber \\
 &=&\log_2 M \:.
\eea
Therefore,
\bea
I(X;Y)=\log_2 M - \sum_{k,k_E}p(k|k_E)\log_2 \frac{1}{p(k|k_E)}\:\:.
\eea

Eq.~(\ref{overlap}) gives the un-normalized Conditional Probability
\bea
p(k|k_E)=e^{-|\alpha|^2\left[
1-\cos \left[
  (\pi/M)  \left( k-k_E)\right]
  \right)
\right]}\:.
\eea
The notation designates an angle set by the legitimate users as $\phi=k \pi/M$ and an angle set by the adversary as
$\phi_E=k_E \pi/M$.
The normalized form of the Conditional Probability will be written
\bea
p_{\tiny \mbox{norm}}(k|k_E)\!\!=\!\!\frac{e^{-|\alpha|^2\left[
1-\cos \left[
  (\pi/M)  \left( k-k_E)\right]
  \right)
\right]}}{\sum_{k=0}^{M-1} \sum_{k_E=0}^{M-1} e^{-|\alpha|^2\left[
1-\cos \left[
  (\pi/M)  \left( k-k_E)\right]
  \right)
\right]}}.
\eea

\begin{figure}[h!]
\centerline{\scalebox{0.25}{\includegraphics{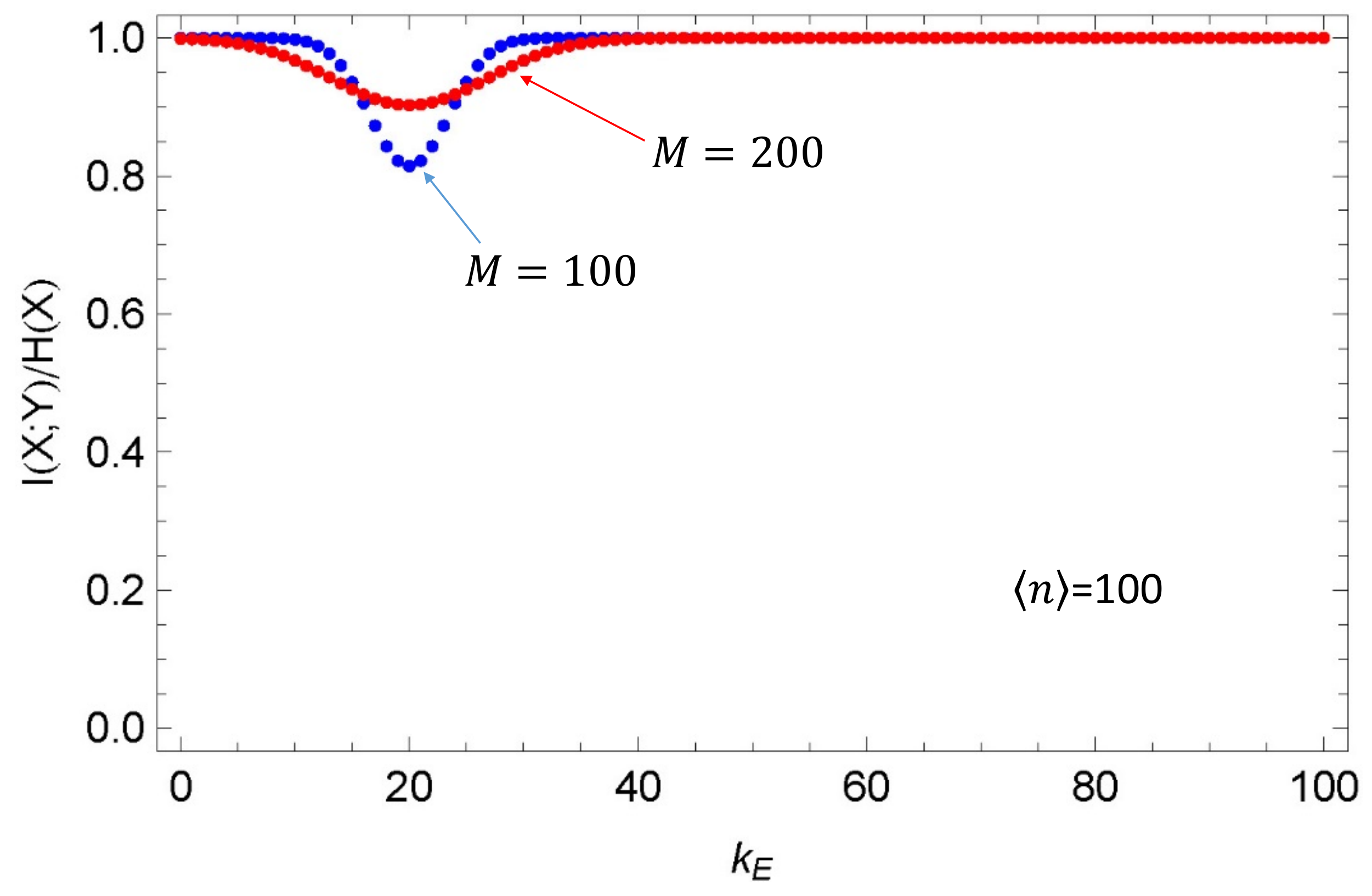}}}
\caption{ \label{IxyoverHx}
Example of the relative information $I(k;k_E)/H(k)$ gained by Eve on $H(k)$ when she learns $k_E$ from her measurement. When $I(k=20;k_E)/H(k=20)\rightarrow 1$ she gains no information on $x$, what occurs for most of the bases values. Only when $k_E$ is set close to $k$ she acquires some information. However, due to the physical noise that produces the physical uncertainty on $k_E$ (or $\phi_E$), {\em even} with $k_E=k$ her knowledge does not give her the desired $I(k=20;k_E)/H(k=20)\rightarrow 0$ but only a limited gain. Her overall probability of success is given by $(1-P_e)$.
 }
\end{figure}

The Mutual Information can be written, in normalized form, for specific values of the phases $\phi$ and $\phi_E$,  using $p_{\tiny \mbox{norm}}(k|k_E)$. One obtains
\bea
&&I(X=\phi;Y=\phi_E)=\frac{1}{M}\log_2 M -\nonumber \\
&&\:\!\!\frac{e^{-|\alpha|^2\left[
1-\cos \left[
  (\pi/M)  \left( k-k_E)\right]
  \right)
\right]}}{\sum_{k=0}^{M-1} \sum_{k_E=0}^{M-1} e^{-|\alpha|^2\left[
1-\cos \left[
  (\pi/M)  \left( k-k_E)\right]
  \right)
\right]}}\nonumber \\
   &&\times \log_2\left[e^{|\alpha|^2\left[
1-\cos \left[
  (\pi/M)  \left( k-k_E)\right]
  \right)
\right]} \right. \nonumber \\ && \times     \left.
\sum_{k=0}^{M-1} \sum_{k_E=0}^{M-1} e^{-|\alpha|^2\left[
1-\cos \left[
  (\pi/M)  \left( k-k_E)\right]
  \right)
\right]}
\right]\:\:.
\eea
One may interpret $I(X=\phi;Y=\phi_E)$ as the average reduction in uncertainty about $\phi$  when Eve in some way learns $\phi_E$. As the $k$ values are uniformly distributed the entropy of $\phi$ (or $k$) is
\bea
H(k)=\frac{1}{M}\log_2 \frac{1}{\frac{1}{M}}=\frac{1}{M}\log_2 M\:,
\eea
The relative reduction in uncertainty obtained by Eve can be quantified by
\bea
r_{I/H}\equiv \frac{I(k;k_E)}{H(k)}\:\:\:\left( 0 \leq r_{I/H} \leq 1 \right)\:\:.
\eea
Fig.~\ref{IxyoverHx} shows $I(k=20;k_E)/H(k=20)$ for $|\alpha|^2=\langle n\rangle =100$ and two set of bases, $M=100$ and $M=200$. The value 20 is arbitrary, as any other value gives similar results.

Going back to the question
{\em ``Will this gained information increases Eve's knowledge on the basis or bit sent?"},
one concludes that she gains some information around the basis value $k$ used by the legitimate users but no knowledge for distant values from $k$.
Therefore, the small amount of information gained by Eve just by excluding bases numbers distant from the basis value set, will be {\em irrelevant} after PA methods reduce this leaked information to a negligible level. Therefore, the amount of Eve's useful information about the bit pool is negligible.

\section{Conclusions}

The key distribution system introduced in \cite{barbosa1} was revisited and improved with inclusion of a specific protocol for the key distribution that includes both the noise protected step and a PA protocol. The PA protocol uses the bit pool shown in Fig.~\ref{AliceBobEveAirGap_23June2015}.

It was shown that starting with a shared sequence of $n_0 \: m$ random bits to form physical bases, A and B can distribute in a secure way an {\em unlimited} number of secure bits generated  by a PhRBG generator.
The overall security of the key distribution depends on signal to noise ratio in the transmitted signals, the number of bases $M$ used and the shuffling produced by the PA protocol.

The key distribution process is as a ``one-time-pad booster'' which allow users to use bit-by-bit encryption for {\em top security level} applications and, at the same time, allows fast encryption of large volumes of information.
When working in fiber-optic channels, the system demands, besides the use of a true physical random bit generator working at high speeds, analog-to-digital, digital-to-analog converters, optical modulators and separate channels to avoid perturbation from ordinary signal channels. In optical channels with signals of mesoscopic intensities, the system presents two layers of protection, physical noise and computational difficulties, such as the one exemplified by PA with universal hash  functions.
The system can also be used with classical signals in generic channels using only the computational difficulty guaranteed by the PA protocol when the protection given by the optical noise is not present (noiseless channels).
When used in optical channels both security levels are present.

\section{Acknowledgements}
We acknowledge the support of Minist\'erio da Ci\^encia, Tecnologia e Inova\c c\~ao (MCTI)-Finep(0276/12)-Fundep(19658)-Comando do Ex\'ercito(DCT)-RENASIC.

\appendix{}

\section{Privacy Amplification - Toeplitz matrices}
\label{Privacy Amplification - Toeplitz matrices}

Privacy Amplification can be used to increase the security level already given by the physical noise in the channel.
The PA procedure establishes \cite{BBCM} that once A have sent to B a sequence $a$ of bits $(n=\mbox{Length}[n_0])$, if Eve obtains an estimated number of $st$ bits, A and B could reduce the amount of Eve's information.

To achieve this end, A and B need to agree on a procedure that results in a shorter secure string $a'$ on which Eve has an exponentially vanishing knowledge. This procedure demands that the number initial bits $n$ has to be reduced by $st$ and even further by a security parameter $\lambda$ (in bits) to guarantee that Eve can obtain at most $1/\ln(2)2^{\lambda}$ bits (see Section IV of \cite{BBCM}) on $a'$. These operations can be performed on the bit pool shown in Fig.~\ref{AliceBobEveAirGap_23June2015}. The procedure to reduce $a$ to $a'$ may use a hash function $\mathcal H$: $\mathcal{H}a=a'$,
where
$\mathcal{H}$ is a matrix with random elements.

Among the several possible PA choices and for $\mathcal H$ - of random elements, one could choose a matrix where all elements are randomly and independently chosen or even constructed with a starting set of randomly chosen elements. This matrix can be renewed at each distribution bit round for maximum security.

Just to explain the PA process with renewed matrix elements for increased security, an example using
a Toeplitz matrix will be presented. A Toeplitz matrix has a simple structure of form
\bea
{\mathcal H}=\left(
               \begin{array}{ccccc}
                 r_1 & c_2 & c_3 & c_4&... \\
                 r_2 & r_1 & c_2 & c_3&... \\
                 r_3 & r_2 & r_1 & c_2&... \\
                 ... & ... & ... & ...&... \\
               \end{array}
             \right)\:,
\eea
where $r_i$ and $c_j$ are binary random digits taken from the PhRBG. The number of columns should be equal to the number of fresh bits $n$ to be transmitted plus the number of bits $n \: m$ (secretely shared by A and B) to generate the modulation bases for the transmitted bits.  The number of rows is equal the $n+n\times m-n t_{\tiny \mbox{leak}}-\lambda$. The number of bits in a column is the same number of bits in $a$. This way, an input bit stream, or a vector with $n(m+1)$ components (bits) gives an output of $n(m+1)-n t_{\tiny \mbox{leak}}-\lambda$ bits, which is the desired reduction in the number of bits for A and B such that Eve has a negligible knowledge on them.

{\em What is the number of secure bits finally available for OTP? --}
What was just described was a PA protocol applied for a first run
starting with $n_0(m+1)$ bits. The output number of bits $n_0$ was reduced to $n_1(m+1)-n_1 t_{\tiny \mbox{leak}}-\lambda$ bits.

As shown in the PA protocol (Section~\ref{PA}), after the first round both Alice and Bob share a distilled sequence of $\overline{s}$ secure bits to be used as OTP and still have a shared fresh sequence of bases bits $s\: m$.

The process is {\em unlimited} in number of runs and will be as fast as the current technology allows because there are no fundamental physical limitations in the bit generation occurring in the PhRBG.

\section{Stoke´s parameters of a noisy field}
\label{noisyPoincare}

The phase modulation specified by Eq.~(\ref{jz wavefunctions})
\bea
|\Psi(\phi) \rangle=e^{-i J_z \phi} |\Psi_0 \rangle
=|\frac{\alpha}{\sqrt{2}} e^{-i \phi/2} \rangle_x   |\frac{\alpha}{\sqrt{2}} e^{i \phi/2} \rangle_y \:,
\eea
is imposed as a phase difference between two orthogonal polarization components represented by annihilation operators $a$ and $b$, representing fields of equal intensity. This form is due to the optical modulator being considered. As the imposed electric field (assumed of weak intensity to avoid non-linear effects) travels along the optical fiber, it undergoes randomized polarization fluctuations in direction caused by several somewhat localized effects that modifies the dielectric constants of the supporting glass medium. These effects include thermal fluctuations, acoustic modes, Mie scattering, mechanical stresses,.... The demodulation system represented at the right in Fig.~\ref{AliceBobEveAirGap_23June2015} subtracts the base modulation effects regardless these random contributions, by operating on two arbitrary polarization components.

One may question if the adversary, Eve, would be able to perform phase measurements close to the emitter, such that these complicating perturbations have not taken an appreciable effect yet. Her goal is to extract precise phase information such that she could {\em resolve} the angular separation $\Delta \phi_1$ between two closest bases $k$ and $k+1$. In general, writing $\Delta \phi$ in terms of base indexes $k,k'$, one has $\Delta \phi=(\pi/M)(k-k')$. The question is ``{\em what is the effect on the inherent optical shot-noise on her measurements?}".  An equivalent but more precise question would be ``{\em what is the maximum resolution on $k-k'$ possible to be achievable by Eve given an average number of photons $\langle n \rangle$ and a number of bases $M$?}"

\begin{figure}[h!]
\centerline{\scalebox{0.4}{\includegraphics{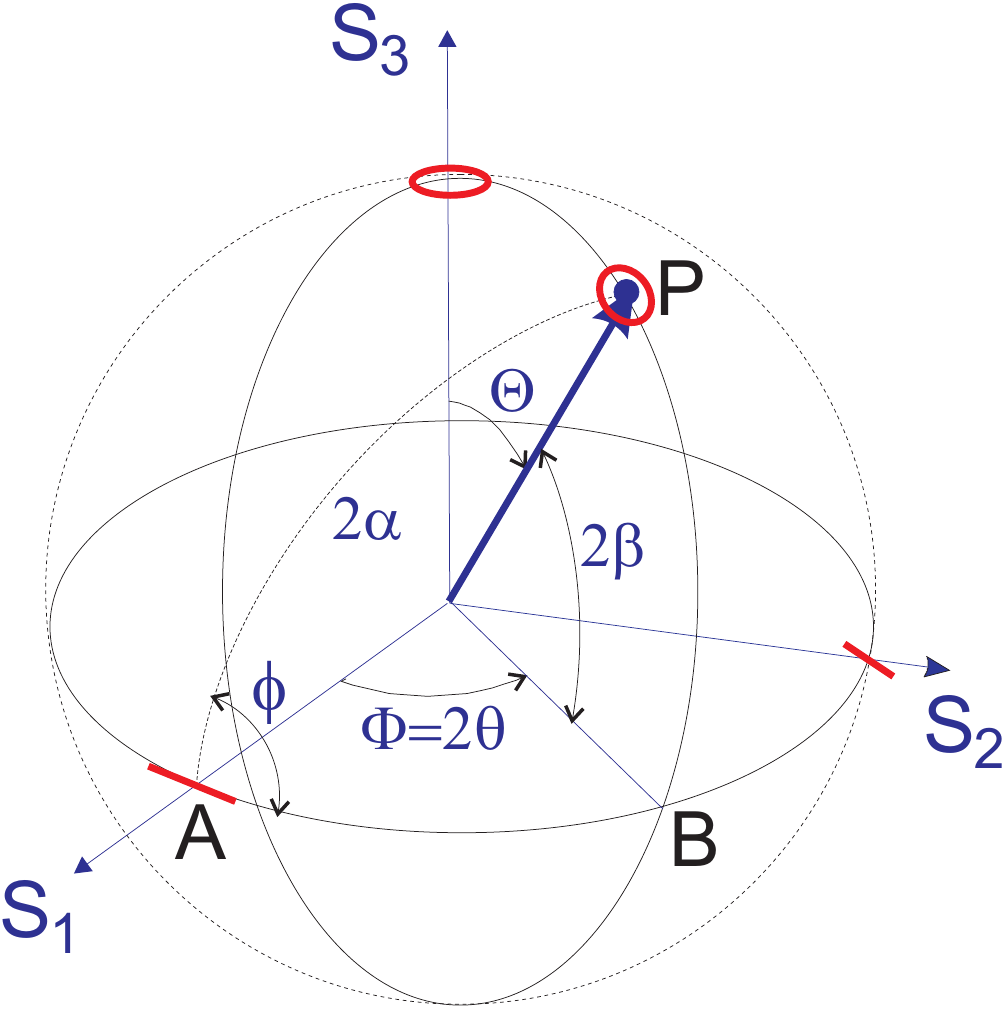}}}
\caption{ \label{Poincare_sphere} $\Theta$ and $\Phi$ are the polar and azimuthal angles that indicate a point P on the sphere. Polarizations on the equator of this sphere represent linearly polarized states with different inclination angles. For example, a point on $\Phi=0,\Theta=\pi/2$ represent a linearly polarized state along $x$, for light propagating along $z$, while its antipodal point $\Phi=\pi,\Theta=\pi/2$ represents a linearly polarized state along $y$. Similarly, a point $\Theta=0$ represents   a (+) circularly polarized light state and point $\Theta=\pi$ a (-) circularly polarized light state. Arbitrary points, like P, represent elliptical polarization states. }
\end{figure}

In order to answer this question, one may start recalling some adequate tools  such as Stokes parameters and the Poincar\'e sphere.
In order to understand transformations of an optical medium or a device on any incoming light mode, it is useful to depict the Poincar\'e sphere of polarizations (See Fig.~\ref{Poincare_sphere}) and to write the incoming polarized electric field in terms of the variables for this sphere.
A polarization state is represented by a point on the Poincar\'e sphere given by the two coordinates $\Theta$ (polar angle) and $\Phi$ (azimuthal angle). The Poincar\'e sphere has its radius defined by the Stokes' parameter $s_0$ (equal to the {\em intensity} of the polarized light) and the projections of the point $(\Theta,\Phi)$ on the orthogonal axes S$_1$, S$_2$ and S$_3$. These projections have values $s_1$, $s_2$ and $s_3$ and are known as Stokes' parameters (Stokes, 1852) \cite{Wolf}. Therefore, the Stokes parameters describe a general polarized light state.

  Physical analyzers such as the crystal axis of a polarization beam splitter or wave-plates produce field projections onto the Poincar\'e's sphere axes  and allow for photon number or intensity measurements, leading to  $s_0,s_1, s_2, s_3$.

Operations and operators describing these measurements are known either in the classical or quantum domain. In the quantum domain these parameters are given by the expectation values of the Hermitian operators of the number operator $\widehat{N}$ and of the total angular momentum of light as described by Schwinger in terms of two bosonic modes given by annihilation operators $\widehat{a}$ and $\widehat{b}$
\bea
\label{Schwinger}
\widehat{s}_0=\widehat{a}^{\dagger}\widehat{a}+\widehat{b}^{\dagger}\widehat{b}&=& \widehat{N}\\
\widehat{s}_1= \widehat{a}^{\dagger}\widehat{a}-\widehat{b}^{\dagger}\widehat{b}&=& 2 \widehat{J}_z=\widehat{\sigma}_z
\\
\widehat{s}_2=\widehat{a}^{\dagger}\widehat{b}+\widehat{b}^{\dagger}\widehat{a}&=& 2 \widehat{J}_x=\widehat{\sigma}_x\\
\label{Schwinger_f}
\widehat{s}_3=\frac{1}{i}(\widehat{a}^{\dagger}\widehat{b}-\widehat{b}^{\dagger}\widehat{a})&=& 2 \widehat{J}_y=\widehat{\sigma}_y\:\:,
\eea
from which
$\langle \widehat{s}_0 \rangle=s_0$, $\langle \widehat{s}_1 \rangle=s_1$, $\langle \widehat{s}_2 \rangle=s_2$, $\langle \widehat{s}_3 \rangle=s_3$.


At this point it is interesting to note that Eqs.~(\ref{Schwinger}) to (\ref{Schwinger_f}) define operators $J_i,\:(i=x,y,z)$ in terms of boson operators $a$ and $b$ (hats will be ignored from now on) and that they obey the same commutation properties as the ones connected with angular momentum:  $\left[  J_i,J_k  \right]=i \epsilon_{ijk}J_k$. This leads to the conservation of the total number of photons $n$ as they go through a lossless optical device:  $n=n_a+n_b=a^{\dagger}a+b^{\dagger}b$.

Standard procedures to perform these measurements have been well established \cite{Wolf} and, from the experimental side, even automated measuring systems can be found to perform these tasks. For example, designating an intensity by $I$ and by $x$ the horizontal axis $H$ and by $y$ the vertical axis $V$, and by the indexes  $R$ and $L$, circular states of light, one could write
\bea
s_0&=&I_H +I_V=a^2+b^2\\
s_1&=&I_H -I_V=a^2-b^2=s_0 \cos(2 \beta) \cos(2 \theta)\\
s_2&=&I_{45^0}-I_{-45^0}=2 a b \cos \phi=s_0 \cos(2 \beta) \sin(2 \theta)\\
s_3&=&I_R-I_L=2 a b \sin \phi=s_0 \sin(2 \beta)\:\:.
\eea
\begin{figure}[h!]
\centerline{\scalebox{0.2}{\includegraphics{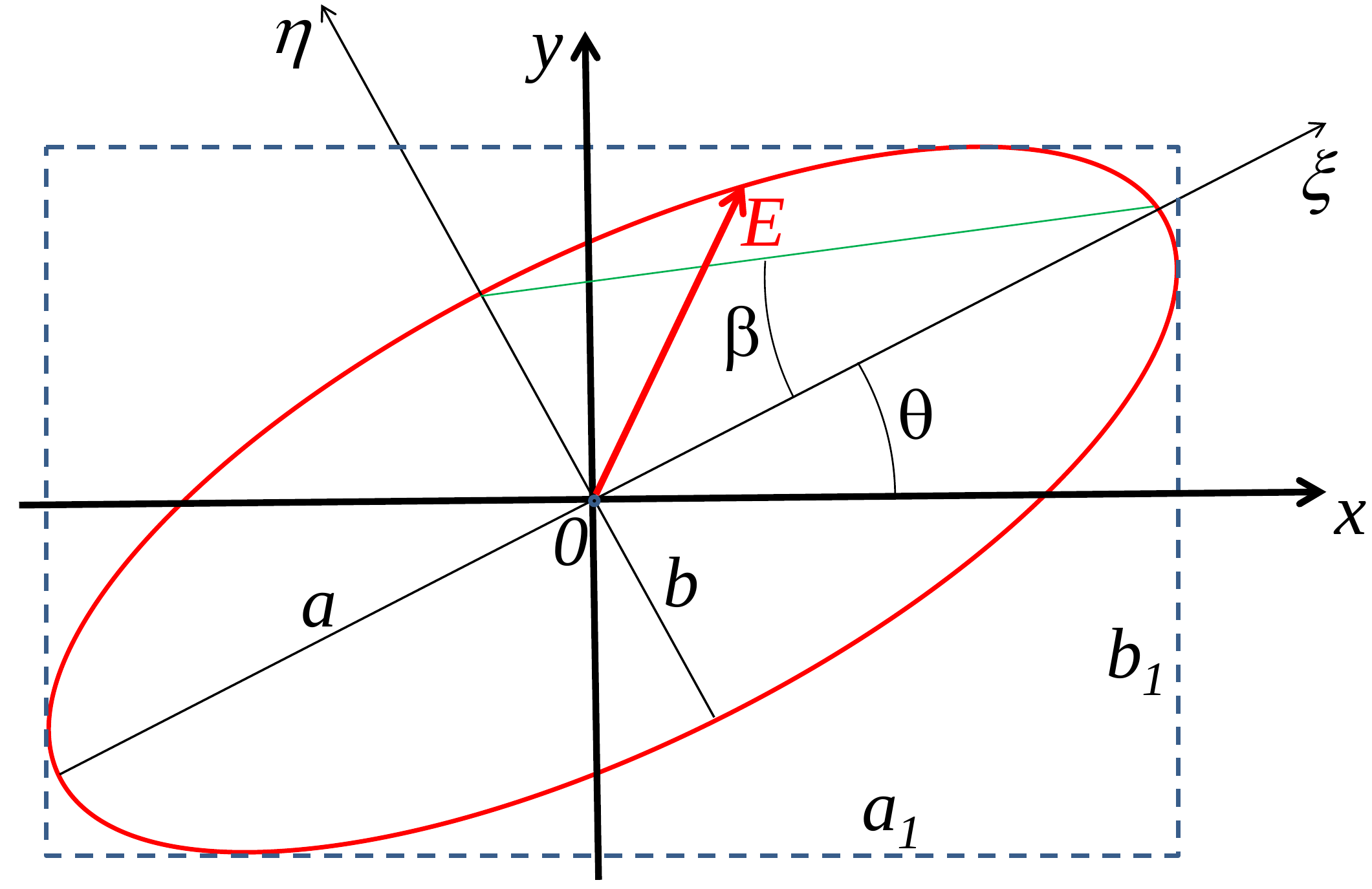}}}
\caption{ \label{Ellipse} A polarization state of light represented by an ellipse, with principal axes along $\xi$ and $\eta$, depicted with the main axes $x$ and $y$. $\beta$ can be either positive or negative, giving the senses in which the ellipse may be described (See Fig.~\ref{Poincare_sphere}, top or bottom half hemisphere).}
\end{figure}
See Fig.~\ref{Ellipse} for definitions.
 Light noise associated to a source of light cannot be eliminated and noise-to-signal ratio cannot be be rendered arbitrarily negligible. The influence of the noise on the propagated signals can be explored in several ways, including for cryptographic purposes.
 The associated error in these measurements are by far less established and belong to the quantum research realm (\cite{carruthers-nieto,clarke-chefles-barnett-riis}). Generalized quantum measurements have been applied to separate deterministically two nonorthogonal quantum states and add a necessary set of inconclusive results \cite{hut-gis}. The subject is a permanent area of research \cite{SotoKHLeuchs}.

 To see the effect of noise, one may calculate
 $\langle \widehat{J}_z \rangle$, $\langle \widehat{J}_x \rangle$, $\langle \widehat{J}_y \rangle$
 and the associated variances
 $\sigma_z^2=\langle \left( \widehat{J}_z-\langle \widehat{J}_z\rangle \right)^2 \rangle$,
 $\sigma_x^2=\langle \left( \widehat{J}_x-\langle \widehat{J}_x\rangle \right)^2 \rangle$,
 $\sigma_y^2=\langle \left( \widehat{J}_y-\langle \widehat{J}_y\rangle \right)^2 \rangle$ .

\begin{figure}[h!]
\centerline{\scalebox{0.3}{\includegraphics{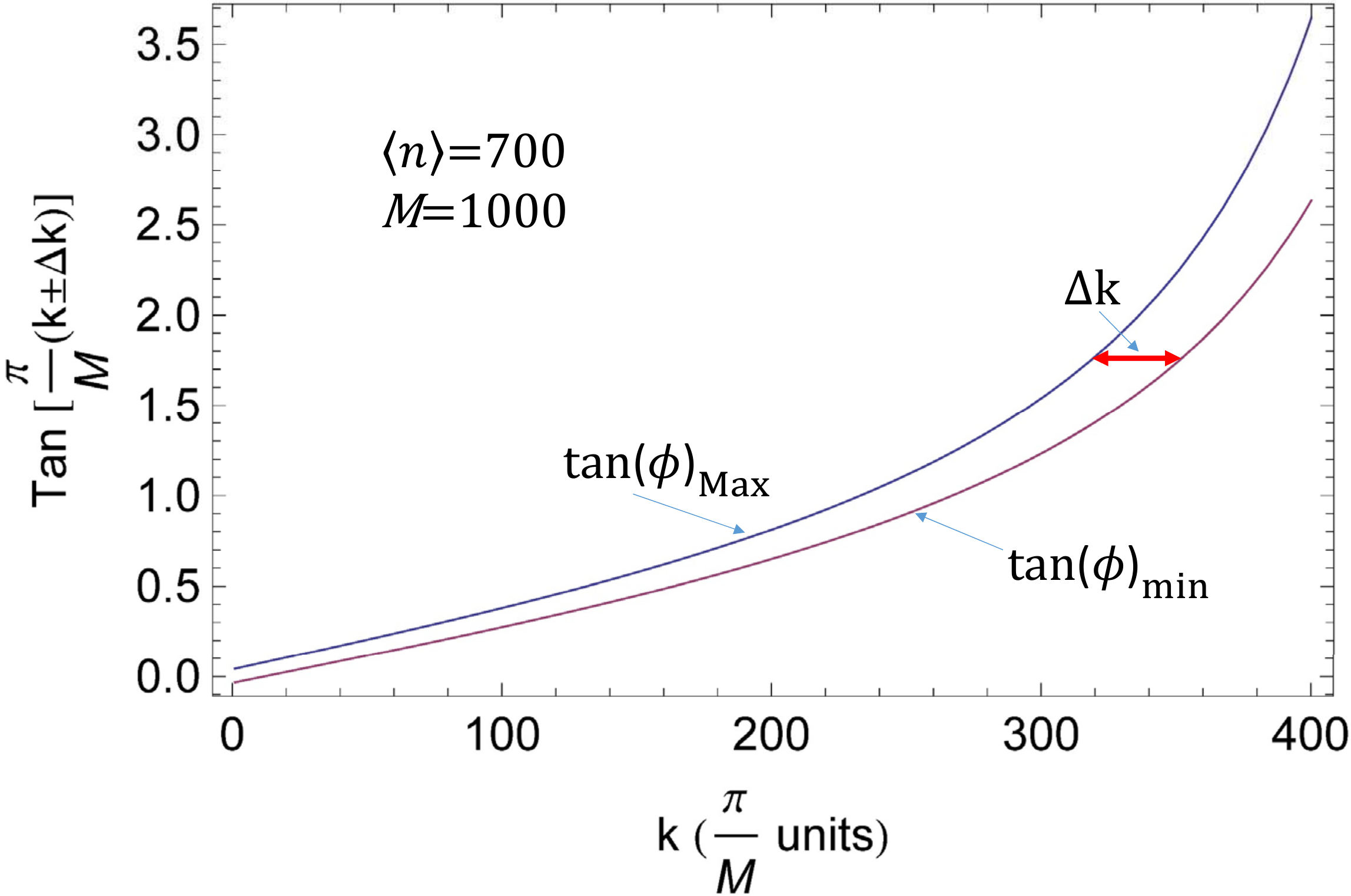}}}
\caption{ \label{pM1000n700}  A sample of extrema for  $\tan\left( \frac{k}{M}\pm \Delta k \right)$ versus $k$.  }
\end{figure}

\begin{figure}[h!]
\centerline{\scalebox{0.27}{\includegraphics{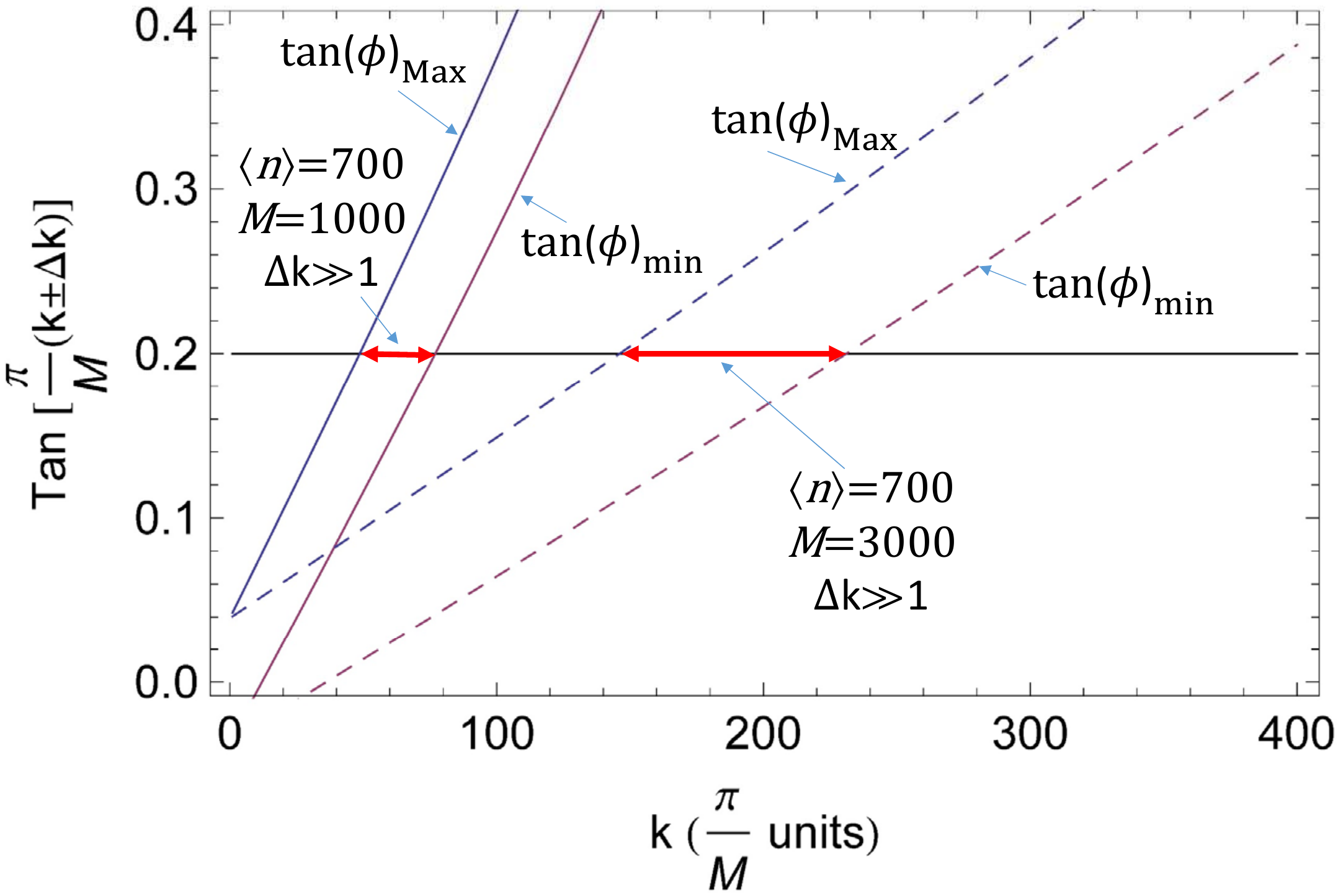}}}
\caption{ \label{Sh} Extrema for  $\tan  \left( \frac{k}{M} \pm \Delta k \right)$ versus $k$.
}
\end{figure}

\begin{figure}[h!]
\centerline{\scalebox{0.27}{\includegraphics{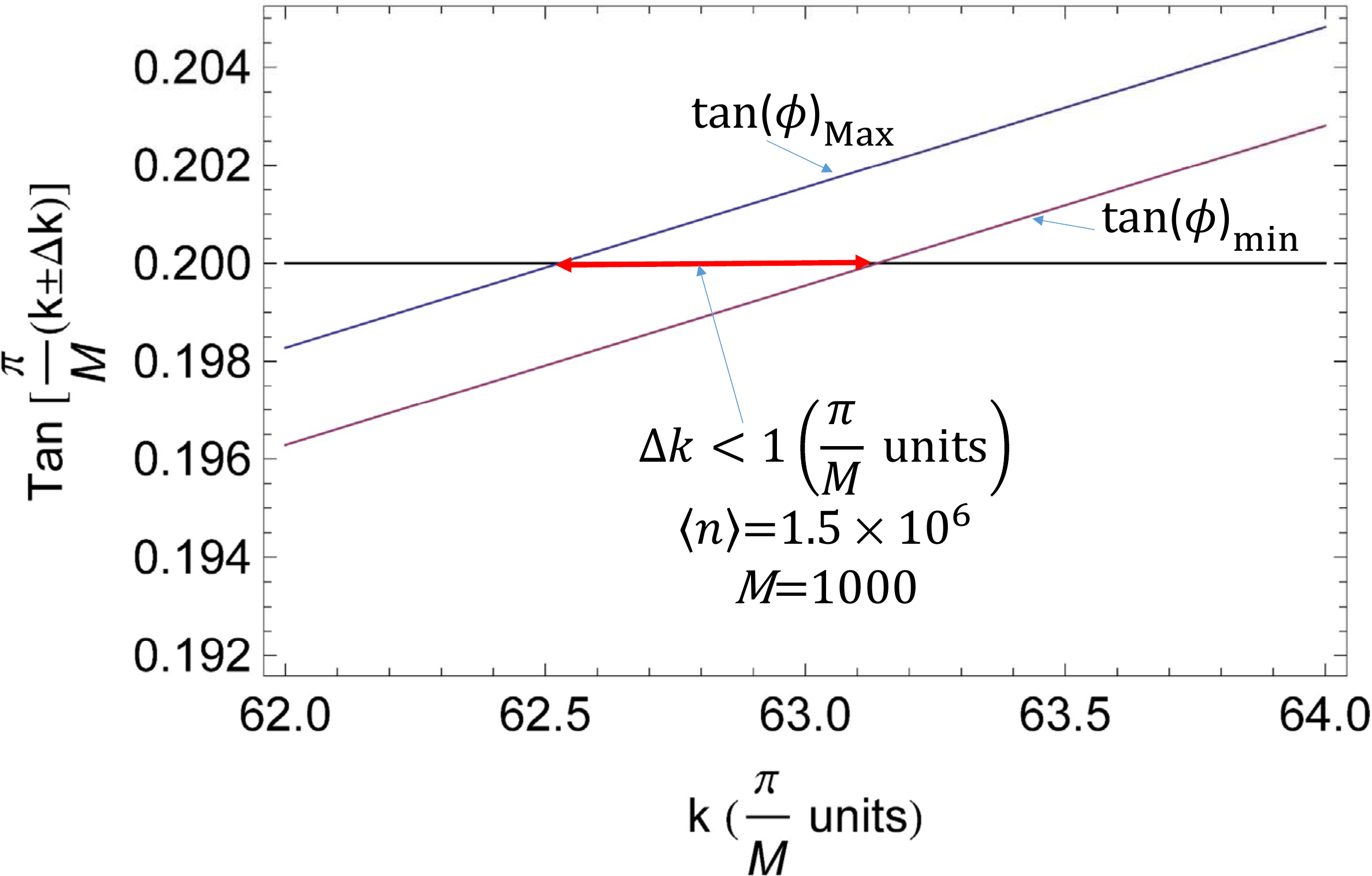}}}
\caption{ \label{sb}   Extrema for  $\tan\left( \frac{k}{M}\pm \Delta k \right)$ versus $k$.
}
\end{figure}

 Observing that $\langle \left(J_i -  \langle J_i \rangle    \right)\left(J_k -\langle J_k \rangle \right)\rangle=\langle J_i J_k \rangle-\langle J_i \rangle \langle J_k \rangle$, the quantities $\langle J_i J_k \rangle$ and $\langle J_i \rangle$ have to be calculated for the $x,y,z$ components. Expansion of the $J_i J_k$ products in normal order and application of the operators on the wave-function given by Eq.~(\ref{jz wavefunctions}) is straightforward. One obtains
\bea
\langle \psi| J_x J_x |  \psi \rangle&=&  \frac{\langle n \rangle}{8}\left[ 2+\langle n \rangle \left(1+ \cos(2 \phi)  \right)\right] \\
\langle \psi| J_x J_y |  \psi \rangle&=&  \frac{\langle n \rangle}{8}\sin(2 \phi)\\
\langle \psi| J_x J_z |  \psi \rangle&=&  -i \frac{\langle n \rangle}{4}\sin\phi\\
\langle \psi| J_y J_x|\psi \rangle &=&  \frac{\langle n \rangle}{8}\sin(2 \phi) \\
 \langle \psi| J_y J_y|\psi \rangle &=&   \frac{\langle n \rangle}{8}\left[ 2+\langle n \rangle \left(1- \cos(2 \phi)  \right)\right] \\
 \langle \psi| J_y J_z|\psi \rangle &=& i \frac{\langle n \rangle}{4}\cos\phi \\
 \langle \psi| J_z J_x|\psi \rangle &=& i \frac{\langle n \rangle}{4}\sin\phi \\
 \langle \psi| J_z J_y|\psi \rangle &=&  -i \frac{\langle n \rangle}{4}\cos\phi \\
 \langle \psi| J_z J_z|\psi \rangle &=&  \langle n \rangle\:,\\
\langle \psi| J_x |  \psi \rangle\!\!=\!\!  \frac{\langle n \rangle}{2} \cos\phi \!\!\!\!\!&&\!\!\!\!\!\!\!\!,
 \langle \psi| J_y|\psi \rangle \!\!=\!\!  \frac{\langle n \rangle}{2} \sin\phi,
 \langle \psi| J_z|\psi \rangle \!\!=\!\! 0.
\eea
The ratio $\langle \psi| J_y|\psi \rangle/\langle \psi| J_x |  \psi \rangle$ of expected values of the Hermitian operators would give a measure of $\tan \phi$ and, therefore, of $\phi$ -- if not for the deviations produced by the light noise. Considering these deviations one has
\bea
\tan\left( \phi\pm \Delta \phi\right)
=\frac{     \langle \psi| J_y|\psi \rangle \pm \sigma_y    }{   \langle \psi| J_x |  \psi \rangle \pm \sigma_x    }
=
\frac{  \sin\left[ \frac{\pi}{M} k\right] \pm \frac{1}{  \sqrt{  \langle n \rangle}} }{ \cos\left[ \frac{\pi}{M} k\right] \pm \frac{1}{  \sqrt{  \langle n \rangle}}  }\:,
\eea
where $\phi$ was written in the discrete set of $M$ $k$ phase values. In order to get the extrema of $\tan\left( \phi\pm \Delta \phi\right)$ one writes:
\bea
\tan \phi_{\mbox{\tiny Max}}=\frac{  \sin\left[ \frac{\pi}{M} k\right] + \frac{1}{  \sqrt{  \langle n \rangle}} }{ \cos\left[ \frac{\pi}{M} k\right] - \frac{1}{  \sqrt{  \langle n \rangle}}  },\:
\label{max}
\tan \phi_{\mbox{\tiny min}}=\frac{  \sin\left[ \frac{\pi}{M} k\right] - \frac{1}{  \sqrt{  \langle n \rangle}} }{ \cos\left[ \frac{\pi}{M} k\right] + \frac{1}{  \sqrt{  \langle n \rangle}}  }
\label{min} \nonumber
\eea
Fig.~\ref{pM1000n700} shows Eqs.~\ref{max} and \ref{min} in a range of values. $\Delta k$ represents the irreducible uncertainty due to the phase noise. In this example $\Delta k \gg 1$.
Fig.~\ref{Sh} and Fig.~\ref{sb} show the $\Delta k$  for a different set of values $\langle n \rangle$ and $M$.
Fig.~\ref{Sh} shows that for a fixed $\langle n \rangle$ the uncertainty $\Delta k$ (or $\Delta \phi$) increases with the number of bases $M$ used.
Fig.~\ref{sb} shows that for an intense field $\langle n \rangle \gg 1$ the uncertainty $\Delta k$ can be reduced giving the resolution $\Delta k \ll 1$ or $\Delta \phi < \pi/M$. In this condition of intense fields, the adversary could identify any basis used and, therefore, obtain the bit sent from A to B. This shows that A and B can frustrate the adversary by choosing $\langle n \rangle$ and $M$ such that the adversary cannot distinguish which basis was used in every emission. The POVM calculation shown in \cite{barbosa1} details this in a complementary and quite general way.

  \begin{biography}[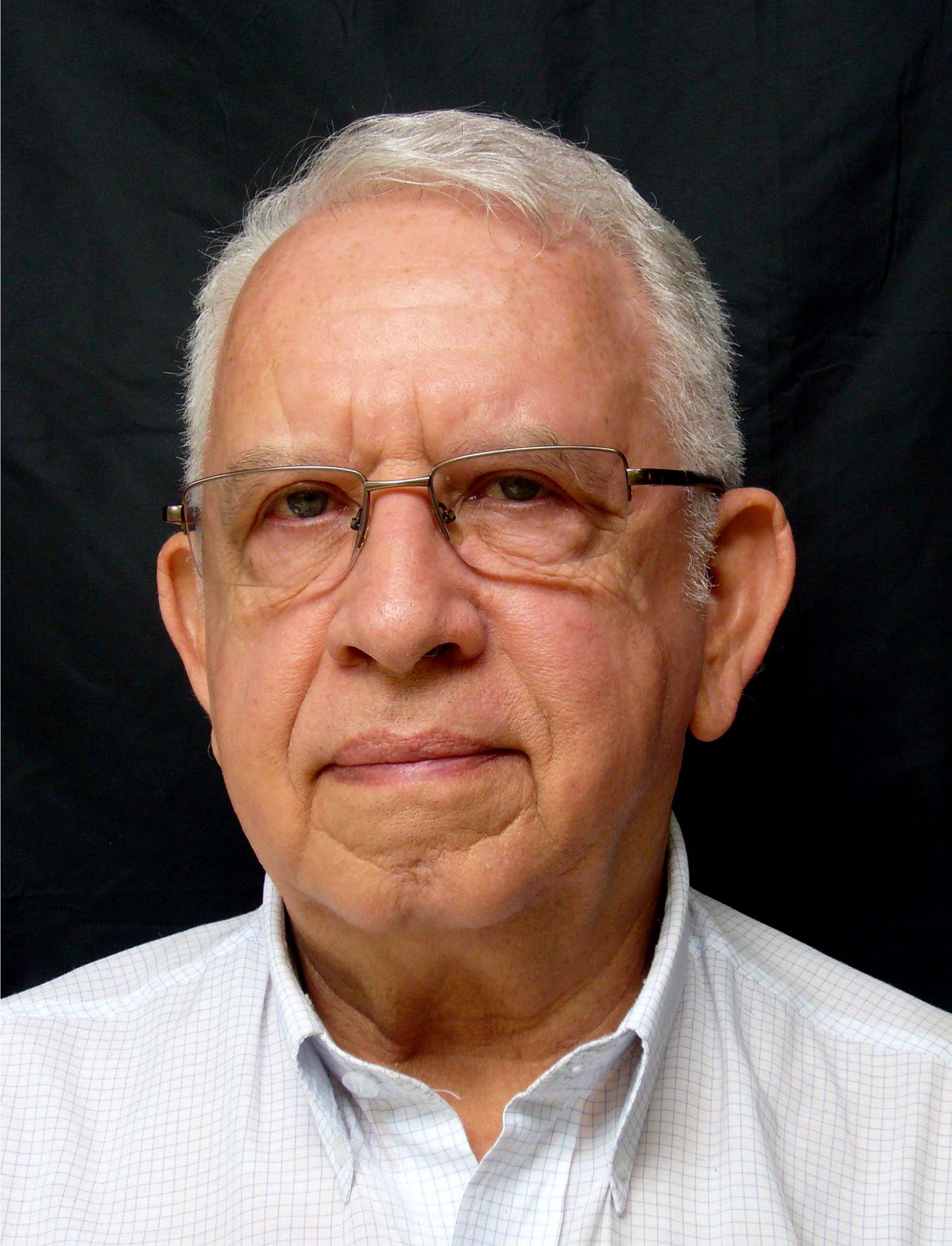]{}{{\bf G. A. Barbosa}-- PhD (Physics)/University of Southern California, 1974. Areas of work: Quantum Optics, Condensed Matter (Theory and Experiment), Physical Cryptography.
Full Professor, Universidade Federal de Minas Gerais/MG/Brazil (up to 1995)/Northwestern University (2000/2012), and CEO, QuantaSec Consultoria e Projetos em Criptografia F\'isica Ltda /Brazil.
}
 \end{biography}
  \begin{biography}[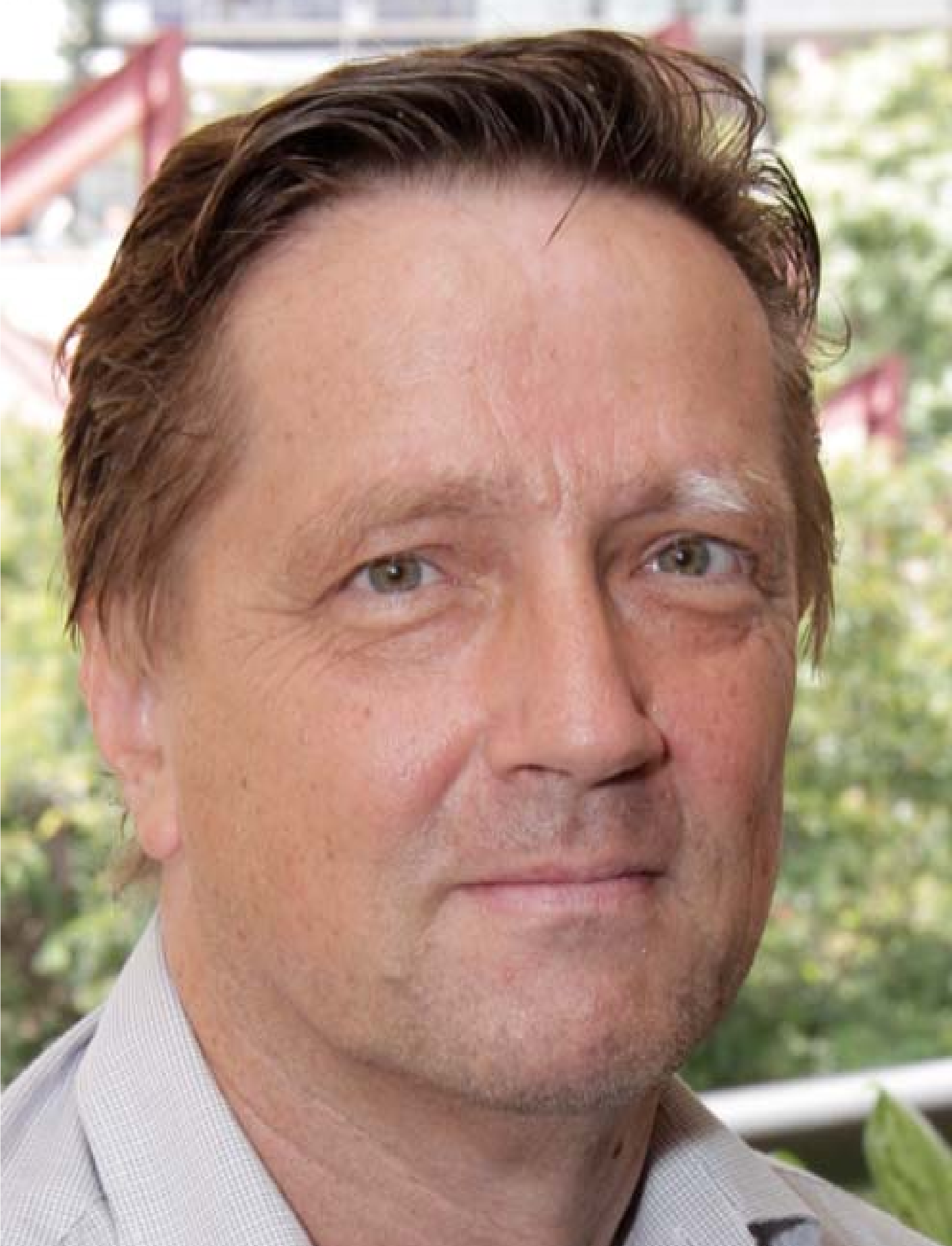]{}{{\bf J. van de Graaf}-- PhD (Informatics)/Universit\'e de Montr\'eal, 1998. Areas of work:
  Cryptography -- theoretical and the applied aspects of cryptographic protocols.
Assistant Professor at the Universidade Federal de Ouro Preto (August 2008/January 2011). Professor  at the Universidade Federal de Minas Gerais (March 2011 up to date).
}
 \end{biography}

\end{document}